\documentclass[fleqn,usenatbib]{mnras}

\usepackage{newtxtext,newtxmath}
\usepackage[T1]{fontenc}

\DeclareRobustCommand{\VAN}[3]{#2}
\let\VANthebibliography\thebibliography
\def\thebibliography{\DeclareRobustCommand{\VAN}[3]{##3}\VANthebibliography}

\usepackage{graphicx}	% Including figure files
\usepackage{amsmath}	% Advanced maths commands

\newcommand{\vect}[1]{\boldsymbol{#1}}
\usepackage{xcolor}
\usepackage{ulem}

\title[Cluster halo shapes in CDM and SIDM models]{Cluster halo shapes in CDM and SIDM models: Unveiling the DM particle nature using a weak lensing approach.}

\author[E. J. Gonzalez et al.]{ 
Elizabeth J. Gonzalez,$^{1,2,3}$\thanks{also at Port d'Informació Científica (PIC), Campus UAB, C. Albareda s/n, 08193 Barcelona (Barcelona), Spain.}\thanks{E-mail: ejgonzalez@unc.edu.ar},
Agustín Rodríguez-Medrano$^{2,3}$,
\newauthor{
Luis Pereyra$^{2,3}$,
Diego García Lambas$^{2,3}$}
\\
$^{1}$Institut de Física d'Altes Energies (IFAE), The Barcelona Institute of Science and Technology, Campus UAB, 08193 Barcelona, Spain.\\
$^{2}$Instituto de Astronomía Teórica y Experimental (IATE-CONICET), Laprida 854, X5000BGR, C\'ordoba, Argentina.\\
$^{3}$ Observatorio Astron\'omico de C\'ordoba, Universidad Nacional de C\'ordoba (OAC-UNC), Laprida 854, X5000BGR, C\'ordoba, Argentina.\\
}

\date{Accepted XXX. Received YYY; in original form ZZZ}

\pubyear{2015}

\begin{document}
\label{firstpage}
\pagerange{\pageref{firstpage}--\pageref{lastpage}}
\maketitle

% Abstract of the paper
\begin{abstract}
Self-interacting dark matter (SIDM) is an alternative to the standard collisionless cold dark matter model (CDM), allowing for interactions between the dark matter particles through the introduction of a self-scattering cross-section. However, the observable effects between these two scenarios are hard to detect. In this work we present a detailed analysis of an application of galaxy-galaxy lensing to measure with high precision the shapes of cluster halos and how this approach can be used to obtain information regarding the nature of the dark matter particle. Using two sets of simulated data, SIDM and CDM simulations, we compute stacked shear maps centred on several subsets of halos with masses $\gtrsim 10^{13.5} M_\odot$. From these maps, we obtain the quadrupole profiles related to the mean projected elongation of the particle distribution from which the shape parameters are derived. Accounting for a radial shape variation, this technique provides an enhancement of the observed differences between the simulated data-sets. In particular, we obtain a higher slope of the power law for the shape-radial relation for the halos identified in the SIDM simulation, which are rounder towards the centre. Also, as approaching to the mean virial radius, the projected semi-axis ratios converge to similar values than in the CDM simulation. Moreover, we account for the impact of the neighbouring mass, where more strongly elongated distributions are found for the halos in the SIDM simulation, indicating that under dark matter self interaction, the large scale structure imprints a more coherent accretion process. %The presented modelling captures this observed feature allowing a more accurate description of the impact of dark matter self-interaction. 
\end{abstract}

% Select between one and six entries from the list of approved keywords.
% Don't make up new ones.

%%%%%%%%%%%%%%%%%%%%%%%%%%%%%%%%%%%%%%%%%%%%%%%%%%

%%%%%%%%%%%%%%%%% BODY OF PAPER %%%%%%%%%%%%%%%%%%

\section{Introduction}

According to the current cosmological paradigm, $\Lambda$CDM, most of the matter content in the Universe is in the form of cold dark matter (CDM), i.e. non-relativistic and collisionless dark-matter (DM) particles. The success of this model is evidenced in accounting for most of the current cosmological observations with great precision \citep[e.g.][]{Weinberg2013,Alam2017,Aghanim2020}. However, after more than five decades after measuring the galactic rotation curves \citep{Rubin1970}, which is one of the most substantial indicators of this kind of matter, very little is known regarding the nature of these particles. In this context, alternative models that postulate other particle characteristics in addition to those presented by the standard model, become relevant and are interesting subjects to be explored. A strategy to address the lack of knowledge of the DM particle nature is to explore the ranges in which the current cosmological model seems to be less accurate. In this sense, current observations point out at halo scales, as those where potential discrepancies with the model arise \citep[e.g. ][]{Weinberg2015,Bullock2017}.

%Self-interacting dark matter (SIDM) is a theoretical candidate of dark matter that interacts with itself through non-gravitational forces, allowing to elastic scattering. 
Self-interacting dark matter (SIDM) is a theoretical candidate of dark matter that allows strong self interactions, i.e. of an amplitude comparable to the strong force.
It was proposed as an explanation for certain observed astrophysical phenomena that cannot be explained by traditional cold dark matter models \citep{Spergel2000}, like the shallowed core of galaxies \citep{Flores1994, Simon2005,Walker2011} and clusters \citep{Sand2004, Newman2009, Umetsu2012}. Phenomenological, SIDM is an attractive alternative to CDM since it modifies the density distribution at halo scales while conserving the distribution at larger radii, thus leaving intact the successful description of the CDM on large scales. An extensive review of the SIDM model and its predictions can be found in \citet{Tulin2018}.

One of the predictions of the SIDM is the modification of the halo shapes. Within the CDM paradigm halos are triaxial structures, while when considering collisions, the orbits of dark matter particles tend to become more isotropic, resulting in an overall rounder mass distribution. The impact of the self-interactions in shaping the halos has been proposed as a test for determining the DM cross-section \citep{Miralda-escude2002,Peter2013,Brinckmann2018,Despali2022}. However, assessing to the halo shapes through observational techniques is challenging, mainly because of projection effects. A widely employed method for constraining halo shapes is strong lensing. Nevertheless, this approach is mostly limited to the inner radial regions, where the impact of baryons is significant and may lead to potential confusion with the imprints attributed to the SIDM.   

In this scenario, weak-lensing offers a useful approach, since this effect is sensitive to the halo outskirts and can be used to constrain the total halo shape. In particular, weak-lensing measurements have been successfully used to measure the halo shapes of galaxy clusters, both using observations and simulated data \citep{Evans2009,Oguri2010,Clampitt2016,Uitert2017,shin2018}. However, given the lower densities in these regions, the imprints of SIDM are expected to be almost indistinguishable from the CDM predictions. This fact plus the projection effects, make this methodology almost useless for distinguishing between the two proposed scenarios \citep{Robertson2023}. 

In this work, we propose and evaluate using weak-lensing techniques to determine the projected shapes of galaxy cluster halos, with the aim of distinguishing between the predictions of the CDM and SIDM models. In particular, we propose to apply weak-lensing stacking techniques, also known as galaxy-galaxy lensing, to estimate the halo shapes. We expect that the combination of the stacking power with the projection of the lensing signal along the main-axis direction of the halo, can enhance the signal-to-noise ratio, thereby enabling a more precise shape estimation.
In order to test the proposed methodology, we produce two sets of simulated data with the same initial conditions: one following the CDM scenario and the other accounting for dark matter self-interaction. Then, we create lensing maps from both simulations by taking into account the projected density distributions of the particles stacked at the centre of each halo. From these maps we compute the lensing radial profiles and fit them to obtain the aligned mean projected elongation of the stacked halos. Finally, we compare the measurements between the two simulations.

The work is organised as follows, in Sec. \ref{sec:data} we describe the produced simulations and how the halos are identified. We detail the shape computation using the particle distribution and the halo characterisation in Sec. \ref{sec:halos}. The formalism of the lensing analysis is presented in Sec. \ref{sec:lensing}. We present the results of the maps produced and the lensing measurements in Sec. \ref{sec:results}. Finally, we discuss our findings and conclude in Sec. \ref{sec:discussion} and \ref{sec:conclussion}.

\section{Data acquisition}
\label{sec:data}
\subsection{Simulated data}
Multiple analyses of massive groups have constrained the cross-section of the dark matter particle to values of $\sigma /m \lesssim 1.25 , \text{cm}^2/\text{g}$ \citep[see, for instance:][]{Randall2008, Jee2014, Wittman2018}. Taking this into account, we produce two simulations with the same initial conditions and two different dark matter particle cross-sections, $\sigma /m=0\,cm^2/g$ and $\sigma /m=1\, cm^2/g$, named CDM and SIDM simulations, respectively. 
Each simulation consists of two combined boxes with $120\, h^{-1}\rm Mpc$ on side and $1024^{3}$ particles, reaching a resolution of $\sim 10^{8}\, h^{-1}\rm M_{\odot}$.
The cosmological parameters used are $\Omega_m=0.3,\,\Omega_{\Lambda}=0.7$ and $\,H_0=70 \,\rm km \,s/Mpc$. The initial conditions were generated with the \textsc{MUSIC} code \citep{Hahn2011} at $z=50$ and the simulations were run with the \textsc{GIZMO} code \citep{Hopkins2015}. 

The implementations of elastics self-interactions in GIZMO are described in detail in \citet{Rocha2013}. In a nutshell, the probability of scattering in a time step is calculated as:
\begin{equation}
    P_{ij} = \frac{\sigma}{m}m_i v_{ij} g_{ij} \delta t .
\end{equation}
In the equation, $\sigma / m$ is the transfer cross section
per unit mass, $m_i$ the mass of the macroparticle, $v_{ij}$ is the relative velocity between these particles and $g_{ij}$ is a number density factor that accounts for the overlap of two particles smoothing kernels.
For each pair of particles with an interaction probability greater than zero, the interaction is determined by drawing a random number.
If interactions occur, the elastic scattering determines the exit velocity. These velocities are calculated as a function of the masses, the relative velocity and the velocity of the mass centre of the two particles. We use the isotropic scattering implementation, so the direction of scattering is chosen randomly. \citep[][]{Meskhidze2022}. 

\subsection{Halo identification and properties}

From both simulated data-sets snapshots at $z=0$, CDM and SIDM, we identify dark matter halos using the \textsc{rockstar} phase-space halo finder code \citep[][]{Behroozi2013}.  In particular, for our analysis, we consider only host halos, i.e. halos not contained within another larger halo. Through this work, we use several halo properties calculated by \textsc{rockstar} code. All of these calculations are performed on the complete set of particles classified as gravitationally bound within each respective halo. The viral radius, denoted as $R_\text{vir}$, is determined through a profile fitting technique and the viral mass, $M_\text{vir}$, corresponds to the mass enclosed within this radius. The position of each halo is established using the innermost particles of the halo, while the halo-offset, referred to as $X_{\text{off}}$, is defined as the spatial separation between the halo location and its centre of mass. This parameter can be used to establish the halo relaxation state (\ref{subsec:halo_def}).

In this study, we investigate halos with masses $10^{13.5} h^{-1} M_\odot < M_\text{vir} < 10^{15} h^{-1} M_\odot $. Taking into account the position centres of the selected halos in the CDM simulation, we select those in the SIDM simulation with a distance difference between the centres lower than $0.5\, R_\text{vir}$ and differences between the masses up to $20\%$, obtaining in total 501 halos in each simulation. This choice was adopted after varying the distance difference and checking the correlation between the halos according to the measured masses. With the adopted cuts, roughly $97\%$ of the halos identified in the considered mass range, have their counterpart in both simulated data-sets. The proximity and mass similarity criteria used in our matching process, ensure robust comparisons between the two simulated scenarios

We are also interested in testing the variations of the halo shapes according to their local environment. With this aim, we characterise the environment of each halo by computing the distance to the $5^\text{th}$ neighbour in each simulation, $R_5$, considering as neighbours those halos with $M_\text{vir} > 10^{13} h^{-1} M_\odot $. This parameter is a common observable proxy to characterise the local density of the environment \citep[e.g.][]{Lackner2012,Muldrew2012,Ching2017,Santucci2023}. We inspect the variations of this parameter with the halo shapes in \ref{subsec:halo_def}.

In this section, we present the characterisation of halo shapes based on the particle distribution. Additionally, we explore the connection between the observed differences in shape within both simulated data-sets and halo properties, while also defining the subsets of halos to be used in the stacking procedure.

\subsection{Individual halo shape estimates}
\label{subsec:hshape}

In order to perform the stacking procedure, it is necessary to first compute the main orientations of the halos before their combination. With this aim, we characterise the shape of the particle distribution for each halo using only the identified bound particles given by \textsc{rockstar}. Since we are interested in the projected shapes, we obtain the projected particle positions by defining a tangential plane perpendicular to the line-of-sight vector pointing towards each halo centre. We consider all three primary axes of each simulation for the projection, thereby increasing the number of stacked halos.

We calculate the orientation and determine the 3D and projected semi-axis ratios of the particle distribution according to the eigenvectors and eigenvalues of the shape tensors. We use a common definition of the shape tensor, the \textit{reduced} tensor:
\begin{equation} \label{eq:ired}
I^{r}_{ij} \equiv   \frac{ \sum_{n}^{N_p} (x_{i n} \, x_{jn}) / r_{n}^2 }{N_p},
\end{equation}
where the summation runs over the $N_p$ bound particles, $x_{in}$, $x_{jn}$  are the coordinates of the $n-$particle, $r_n$ is the distance to the halo centre and the sub-indexes $i, j\in[0,1,2]$ and $i, j\in[0,1]$ for the 3D and projected positions, respectively. We choose to use the \textit{reduced} tensor definition since it accounts for the radial distance, being more sensitive to the inner particle distribution than the other commonly used definition, the \textit{standard} one \citep[in which $r_n$ is neglected, Eq. 5 in ][]{Brinckmann2018}. Given that the particle density is higher towards the halo centre, the orbits of these particles are more affected by differences in the cross-section. Therefore, a shape definition based on this tensor highlights the differences between the considered simulations.

\begin{figure*}
    \includegraphics[scale=0.4]{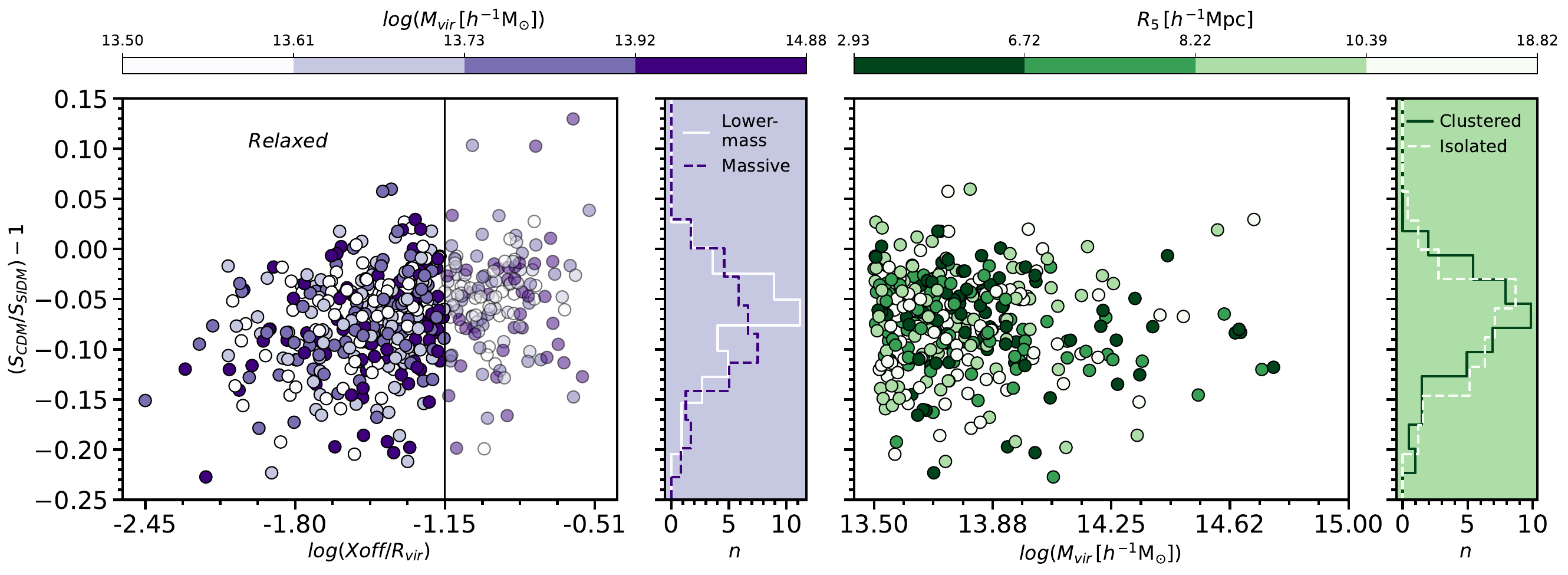}
    \caption{Relation between the halo properties and the shape differences observed in the CDM and SIDM simulations. $S_\text{CDM}$ and $S_\text{SIDM}$ refers to the halos sphericities ($C/A$) measured in the CDM and SIDM simulations, respectively.  In the left panel, we show the relation between $S_\text{CDM}/S_\text{SIDM} - 1$ and the halo relaxation state characterised using $X_\text{off}/R_\text{vir}$. The vertical line indicates the criteria used to select relaxed halos ($X_\text{off}/R_\text{vir} < 0.07$), those halos selected as relaxed are at the left of the solid line in darker colours. The colour code indicates the halo virial masses. In the next panel, we show the normalised $S_\text{CDM}/S_\text{SIDM} - 1$ distributions of the selected halo subsets, \textit{Lower-mass} and \textit{Massive} (see Table \ref{tab:sampdef}). The next panel shows the relation between the shape differences and halos masses. The colour code indicates the $R_5$ parameter, lower values are related to halos in denser environments. Finally, in the last panel, we show the normalised $S_\text{CDM}/S_\text{SIDM} - 1$ distributions for the \textit{Clustered} and \textit{Isolated} selected subsamples (see Table \ref{tab:sampdef}).}
    \label{fig:masa_vecinos}
\end{figure*}
\section{Halo characterisation}
\label{sec:halos}

By diagonalising the tensors, we obtain the semi-axis values
($A > B > C$ in 3D and $a > b$ in projection) corresponding to the square root of the eigenvalues, while their respective eigenvectors ($\vect{\hat{e}_A}$, $\vect{\hat{e}_B}$ and $\vect{\hat{e}_C}$ in 3D and $\vect{\hat{e}_a}$, $\vect{\hat{e}_b}$ in projection) define the principal axis directions associated with their eigenvalues. From these quantities, we can obtain the semi-axis ratio of the projected particle distribution $q = b/a$ and the position angle of the main semi-axis, $\phi$, computed from $\vect{\hat{e}_a}$. We also obtain for the 3D inertia tensor the sphericity $S = C/A$, which would be used to inspect the differences in the elongation related to the halo properties (see \ref{subsec:halo_def}).

Additionally, we consider an alternative definition of the shape
\citep[see e.g.][]{Katz1991, DubinskiCarlberg1991, Zemp2011}
calculated by using all the particles contained within an ellipsoidal radius ($r_\mathrm{ell}$) given by:
\begin{equation}
r_\mathrm{ell} = \sqrt{x_\mathrm{ell}^2 + \frac{y_\mathrm{ell}^2}{(B/A)^2} + \frac{z_\mathrm{ell}^2}{(C/A)^2} }.
\label{eq:rell3d}
\end{equation}
$x_\mathrm{ell} = \vect{x} . \vect{\hat{e}_A}$, $y_\mathrm{ell} = \vect{x} . \vect{\hat{e}_B}$ and $z_\mathrm{ell} = \vect{x} . \vect{\hat{e}_C}$, where $\vect{x}$ is the position vector of the particle in 3D with respect to the halo centre. Analogously, in 2D, we take into account the particles within an ellipse with radius:
\begin{equation}
r_\mathrm{ell} = \sqrt{x_\mathrm{ell}^2 + \frac{y_\mathrm{ell}^2}{(b/a)^2} }.
\label{eq:rell}
\end{equation}
The procedure for calculating this shape is iterative because the value of the ellipsoidal radius is not known in advance and depends on the semi-axes of the ellipsoid, which need to be determined during the process. In practice, we first consider all the bound particles and calculate the initial set of eigenvectors and eigenvalues. Then, we compute the new set of shape parameters excluding those particles outside the obtained ellipsoidal/ellipse region. This calculation is repeated considering 10 iterations or until the difference between the newly and the previously computed shape parameter ($S$ in 3D or $q$ in projection) represents less than the $1\%$. 

We obtain the shape parameters using both iterative and non-iterative approaches. In general, the semi-axis ratios obtained iteratively exhibit a strong correlation with those calculated using the non-iterative approach, with the latter being approximately $\sim 20\%$ lower. For the stacking procedure, we use the orientations given by the non-iterative method. This last choice is motivated by the fact that in observational studies the distribution of the galaxy members is commonly used as tracers of the underlying dark matter distribution, in order to estimate the cluster orientations \citep{huang2016,Uitert2017,shin2018,Gonzalez2021}. 
 Therefore, due to the typically small number of identified members ($\lesssim 100$), discarding the tracers in each iteration can introduce bias into the shape parameters. This is because the introduced shot noise tends to predict more elongated shapes.
 
 On the other hand, to characterise the individual halo shapes based on the dark matter particle distribution, we use the semi-axis ratios obtained from the iterative method. This is done since this methodology is more commonly used given that leads to less biased results \citep{Behroozi2013,Robertson2023}. Shape definition is a matter of importance specially for this kind of analysis, since halo shapes are particularly difficult to estimate in cored density profiles that
get monotonically more spherical towards the centre, which are expected mainly in the SIDM scenario. A detailed discussion on several shape definitions can be found in \citet{Zemp2011} and in particular for the comparison between SIDM and CDM halo shapes in \citet{Peter2013,Brinckmann2018}. It is important to highlight that these shape parameters are only used for comparative purposes and do not impact the lensing determinations, which constitute the focus of this work.

\subsection{Halo subset definitions}
\label{subsec:halo_def}
The stacking procedure is commonly applied by combining the halos that share a common property, such as a mass proxy. In order to take advantage of this combination technique, we initially investigate how the nature of the particles influences the shaping of cluster halos based on certain intrinsic properties. The properties inspected are the halo relaxation state according to the scaled central position offset, $X_\text{off}/R_\text{vir}$, the halo mass, $M_\text{vir}$ and the halo environment according to $R_5$. 
We characterise the shape differences between the two simulated data sets using the computed 3D semi-axis ratios and the sphericity $S$ according to the iterative method presented in the last section.

In Fig. \ref{fig:masa_vecinos} we show how the ratio between the shape estimates in the SIDM and CDM simulations ($S_\text{CDM}/S_\text{SIDM} - 1$), are related with the considered halo properties. As it can be noticed, relaxed halos classified as those with $X_\text{off}/R_\text{vir} < 0.07$ following \citet{Neto2007}, tend to show larger differences in the shapes. This can be related with the fact that less-relaxed halos yield less precise shape measurements due to a wrong centre definition.  However, we do not obtain a clear relation between the $S_\text{CDM}/S_\text{SIDM} - 1$ and the halos masses or the environment characterisation, $R_5$.

In view of these results, we consider six halo subsets for the stacked analysis, classified considering their relaxation state, mass and environment. Although we do not obtain a clear relation between the shape differences and the environment, we decide to inspect samples selected based on $R_5$ in order to assess the influence of the neighbouring mass component on the analysis. In Table \ref{tab:sampdef} we show the criteria used to define each sample and the number of halos included. Since the analysis is performed by projecting the particles in the three main directions of the simulations, these numbers are tripled in the stacking. In particular, we focus our analysis on relaxed halos since their shapes are better constrained and they are also less affected by systematic effects, such as the centre definition. The samples selected according to the mass and environment are obtained according to the quartiles of the distributions of $M_\text{vir}$ and $R_5$ to enhance the differences. 

\begin{table} 
 \caption{Criteria chosen to select the halo subsets for the stacked analysis and the number of halos included in each sub-sample.}
    \centering
    \begin{tabular}{l l l}
    \hline
    \hline
Sub-sample   & Selection criteria & Number\\
& &  of halos \\
\hline
Total & - & 501 \\
Relaxed & $X_\text{off}/R_\text{vir} < 0.07 $ & 357\\
Isolated & $X_\text{off}/R_\text{vir} < 0.07, R_5 > 10.39 h^{-1}$Mpc & 88 \\
Clustered & $X_\text{off}/R_\text{vir} < 0.07, R_5 < 6.72 h^{-1}$Mpc & 85 \\
Massive & $X_\text{off}/R_\text{vir} < 0.07, M_\text{vir} > 10^{13.92} h^{-1} M_\odot$ & 84 \\
Lower mass & $X_\text{off}/R_\text{vir} < 0.07, M_\text{vir} < 10^{13.61} h^{-1} M_\odot$ & 88 \\
\hline
\end{tabular}
\begin{flushleft}
\end{flushleft}
    \label{tab:sampdef}
\end{table}
\begin{figure*}
    \includegraphics[scale=0.6]{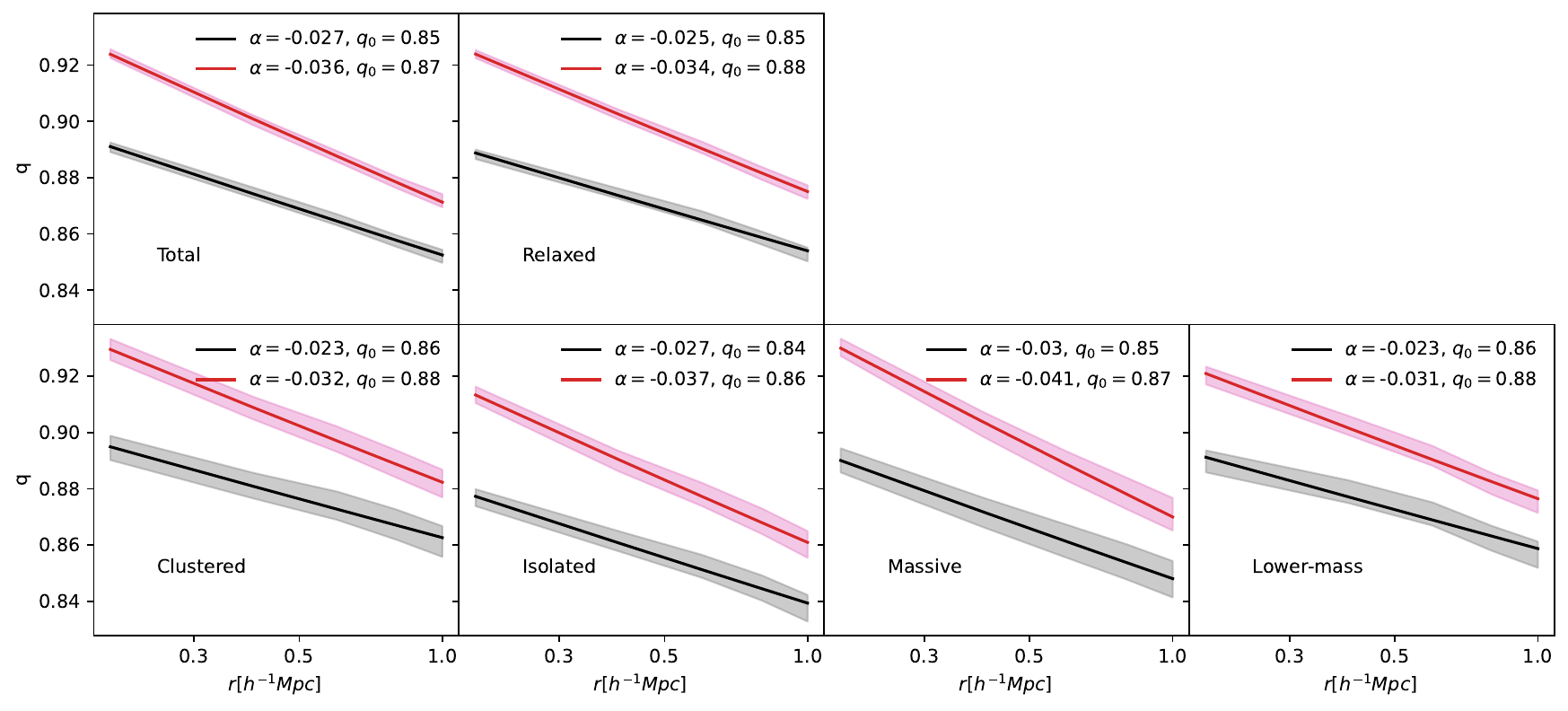}
    \caption{Measured semi-axis ratios according to the stacked particle distribution within a radius $r$, using the \textit{reduced} tensor (see Eq. 2). The shadow regions corresponds to the errors set by a bootstrap with 50 iterations. Grey and pink correspond to the results for CDM and SIDM simulations, respectively. Fitted relations according to Eq. \ref{eq:qr} are shown in red and black and the fitted parameters are shown in the legend.}
    \label{fig:radial_stack}
\end{figure*}
\subsection{Stacked halo shapes}
\label{subsec:stacked_shapes}
For a more direct comparison between the lensing estimates, we perform a stacked analysis of the particle distribution. In this approach, we only consider the bound particles since we intend to characterise the halo shapes. With this aim, we compute the projected shapes of the combined particle distribution for the different halo subsets defined in Table \ref{tab:sampdef}, previously rotating them according to the position angle, $\phi$. We characterise the shapes of the combined halos by computing the shape tensor defined in Eq. \ref{eq:ired}, taking into account the positions of the rotated particles. We perform this procedure by considering the particles within varying radii to get an insight into the shape radial variation. Results are shown in Fig. \ref{fig:radial_stack}. We highlight that the stacking procedure dilutes the scatter introduced by the shape variance in the halo sample and represents the shape radial variation of the mean particle distribution, instead of the mean shapes of the individual halos. Therefore, the shaded areas obtained according to the bootstrap resampling, do not represent the variance of the $q$ distribution, for which we measure a standard deviation of $\sim 0.12$ when considering all bound particles.

The observed radial variation of the semi-axis ratios can be approximately modelled using a power law relation:
\begin{equation} \label{eq:qr}
    q(r) = q_0 r^\alpha.
\end{equation}
We show in Fig. \ref{fig:radial_stack} the fitted parameters, $q_0$ and $\alpha$, for each halo subset. As can be noticed, halos in SIDM simulations tend to be more spherical across the entire range of radii considered. Halos in both simulations tend to be rounder towards the centre, especially those in the SIDM simulation. This is evidenced by the fitted slopes, $\alpha$, being in general $\sim20\%$ larger for the halos identified in the SIDM simulation. 

\section{Halo shape estimates using lensing}
\label{sec:lensing}
%The gravitational lensing effect is originated under the presence of a gravitational field, in this case, generated by the galaxy cluster. This effect bundles the light rays of the luminous sources located behind the gravitational field. This results in the distortion and magnification of the shapes of the background galaxies, i.e. those galaxies that are located behind the cluster.

The gravitational lensing effect is originated under the presence of a gravitational field which bundles the light rays of the luminous sources located behind it. In our case of interest, the gravitational field is generated mainly by the halo which hosts the galaxy cluster and the lensing effect magnifies and distorts the shapes of the background galaxies, i.e. those located behind the cluster. In the inner regions of the clusters, which correspond to the strong lensing regime, distortions are significant leading to the detection of arcs and multiple images. Conversely, in the weak lensing regime at the outskirts of the cluster, the produced effect is smaller and can be only assessed using an statistical approach. The magnitude of the distortion is related to the projected surface density distribution, $\Sigma$, and encoded in the \textit{shear} parameter, $\mathbf{\gamma}$, which is a complex quantity that can be obtained from the ellipticity components of the background galaxies. %In turn, the introduced magnification is related with the \textit{convergence} scalar parameter, $\kappa$. Magnification changes the distribution of background galaxy sizes and their luminosity \citep{Schmidt2012,Huff2014}, modifying the source number densities \citep{Scranton2005,Hildebrandt2009,Hildebrandt2011,Morrison2012,Ford2012,Hildebrandt2013}. The convergence $\kappa$ can be also estimated by measuring the lensing effect in the cosmic microwave background which has the advantage of being less affected by other systematics, such as shape modelling errors in recovering the galaxy shapes, biases in the photometric redshift estimates of background galaxies and contamination by unlensed cluster galaxies \citep[e.g. ][]{Jarvis2016,Melchior2017}.

A key limitation in weak-lensing studies is the relatively low signal-to-noise ratio of the measured effect. Moreover, the observed distortions are combined with the galaxy intrinsic projected elongation, $e_s$. Hence, the observed galaxy ellipticity is a combination of both quantities, $\mathbf{e} \sim \mathbf{e_s} + \mathbf{\gamma}$. The strategy that can be applied to isolate the lensing effect from the measured galaxy ellipticity, is to combine the measurements of many background galaxies located at roughly the same radial distance from the cluster centre. Then, by averaging the tangential ellipticity component of these galaxies, the \textit{shear} estimator can be obtained as $\tilde
{\gamma} = \langle \mathbf{e} \rangle$. 
%In a similar way, this can be applied to increase the signal of the magnification effect to derive $\kappa$. 
This approach assumes that the cluster is spherically symmetric so that the galaxies at the same radial distance are similarly affected by the lensing effect. Also, it assumes that the intrinsic orientations of background galaxies are random, which is valid when the galaxies combined are located in a wide redshift range. With this approach, the precision in the derived lensing estimates will depend on the number of galaxies combined. 

The signal-to-noise ratio can be upgraded by combining the background galaxies of many galaxy clusters that share a similar observed property, such as the number of identified galaxy members or their X-ray luminosity. This strategy is commonly known as stacking techniques or galaxy-galaxy lensing. In this approach, the measured radial \textit{shear}, is related to the mean surface density distribution of the combined clusters \citep[e.g. ][]{Niemiec2017,McClintock2019,Pereira2020}. In principle, this combination results in softening the surface distribution, by blurring the impact of substructure.
However, if the main direction of the projected mass of each cluster is taken into account in the stacking procedure, this technique can be successfully applied in order to obtain the mean aligned projected elongation of the combined clusters \citep{Clampitt2016,Uitert2017,shin2018,Gonzalez2021}. 

%The lensing effect is sensitive to the shape of the surface mass distribution producing an azimuthal variation of the \textit{shear} that can be linked with the projected elongation. Indeed, the mean elongation of the surface mass distribution of a cluster sample can be estimated if the main direction of the projected mass of each cluster is taken into account in the stacking procedure. This approach has been successfully applied to estimate the mean elongation of galaxy clusters. 

In order to perform our analysis, we construct the lensing maps by stacking the particles related with the halos included in the different subsets defined in \ref{subsec:halo_def}, previously rotating them according to the main axis direction of each halo, $\phi$. To take into account that the lensing effect is sensitive to the whole particle distribution along the line-of-sight, all particles within a box of side-length $20 h^{-1}$\,Mpc centred at each halo's position are taken into account. From the lensing maps we compute the radial profiles that can be modeled by considering three halo properties: mass, concentration and shape. In this section, we first summarise the formalism related to the lensing analysis and how the mean ellipticity of combined halos can be obtained using weak-lensing stacking techniques. Then, we describe the computation of lensing estimators for combined halo samples and the derivation of mean halo elongation using this procedure. 

\subsection{Lensing formalism for an elongated mass distribution}

\subsubsection{Surface mass distribution and convergence}

An elliptical surface mass density distribution can be modelled considering confocal elliptical isodensity
contours, $\Sigma(R)$, where $R$ is the elliptical radial coordinate, $R^2 = r^2(q \cos^2(\theta) + \sin^2(\theta)/q)$ \citep{Uitert2017}.  This distribution can be approximated using a multipole expansion in terms of the ellipticity defined as $\epsilon:= (1-q)/(1+q)$ \citep{Schneider1991}:
\begin{equation}
\label{eq:Smyq}
\Sigma(R) = \Sigma(r,\theta) := \Sigma_0(r) + \epsilon \Sigma_2(r) \cos(2\theta) ,
\end{equation}
where we neglect the higher order terms in $\epsilon$. In this approximation, $\theta$ is the angle relative to the major semi-axis of the surface density distribution. $\Sigma_0$ and $\Sigma_2$ are the monopole and quadrupole components, respectively. $\Sigma_0$ is related to the axis-symmetrical mass distribution while the quadrupole component is defined in terms of the monopole as $\Sigma_2 = -r d(\Sigma_0(r))/dr$. 

The surface density distribution can be related with the lensing \textit{convergence} parameter:
\begin{equation}\label{eq:kappa}
    \kappa = \Sigma/\Sigma_{\rm{crit}},
\end{equation}
where $\Sigma_{\rm{crit}}$ is the critical density defined as:
\begin{equation} \label{eq:sig_crit}
\Sigma_{\rm{crit}} = \dfrac{c^{2}}{4 \pi G} \dfrac{D_\text{OS}}{D_\text{OL} D_\text{LS}}.
\end{equation}
Here $D_\text{OL}$, $D_\text{OS}$ and $D_\text{LS}$ are  the angular diameter distances from the observer to the lens, from the observer to the source and from the lens to the source, respectively.

In this work, we adopt a similar approach as presented in \citet{Gonzalez2022} and model the radial surface density distribution, $\Sigma_0$, by taking into account two surface density components: the main halo host and the neighbouring mass contribution. This model assumes that the density distribution of the main halo ($1-$halo term) as well as the contribution of the neighbouring masses ($2-$halo term) are elongated by different amounts and that the main directions are correlated. Thus, although the direction taken into account to align the halos is related with the main halo component, the contribution of the neighbouring mass distribution to the quadrupole is also considered in this modelling. The expected misalignment between the two components will bias the estimated elongation of the neighbouring mass to lower values.

The main halo component is modelled assuming a spherically symmetric NFW profile \citep{Navarro97}, which can be parametrised by the radius that encloses a mean density equal to 200 times the critical density of the Universe, $R_{200}$, and a dimensionless concentration parameter, $c_{200}$. The 3D density profile is given by:
\begin{equation} \label{eq:nfw}
\rho_{1h}(r) =  \dfrac{\rho_{\rm crit} \delta_{c}}{(r/r_{s})(1+r/r_{s})^{2}},
\end{equation}
where  $r_{s}$ is the scale radius, $r_{s} = R_{200}/c_{200}$, $\rho_{\rm crit}$ is the critical density of the Universe and
$\delta_{c}$ is the cha\-rac\-te\-ris\-tic overdensity:
\begin{equation}
\delta_{c} = \frac{200}{3} \dfrac{c_{200}^{3}}{\ln(1+c_{200})-c_{200}/(1+c_{200})}.  
\end{equation}
We define $M_{200}$ as the mass within $R_{200}$ that can be obtained as \mbox{$M_{200}=200\,\rho_{\rm crit} (4/3) \pi\,R_{200}^{3}$}. 
\begin{figure*}     
\includegraphics[scale=0.5]{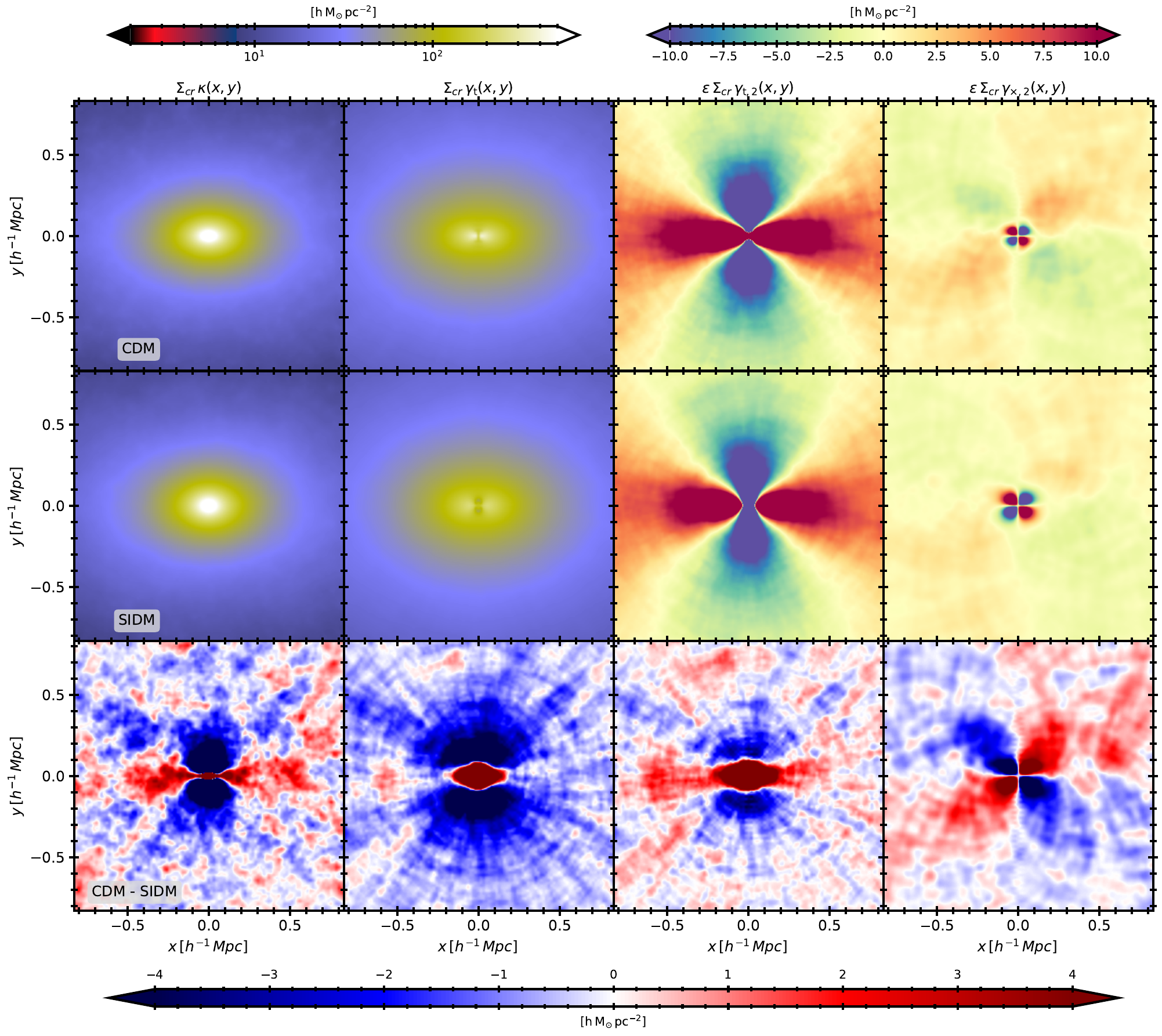}
\caption{Maps of the stacked halos in the CDM (upper panels) and SIDM (middle panels) simulations, and their differences (lower panel). From left to right: The first map corresponds to the surface density map ($\Sigma(x,y) = \Sigma_{cr} \kappa$), computed by stacking the particle distribution of the total sample of halos, taking into account all the particles included within a box of side-length 20$h^{-1}$\,Mpc and after rotating the particles to get the halo major semi-axis aligned with the $x-$axis. The particle positions are centred at each halo position and rotated according to the main axis direction of the halo. The second map is related to the tangential shear, $\Sigma_{cr} \gamma_t(x,y)$, and leads to contrast density profile, $\Delta \Sigma$, when averaged in annular bins. The third map corresponds to the tangential quadrupole component obtained after subtracting the contrast density distribution, $\Delta \Sigma(x,y)$, to the tangential shear map ($\epsilon \Sigma_{cr} \gamma_{t,2} (x,y)$). $\Delta \Sigma(x,y)$ is computed with the fitted mass and concentration derived from adopted modelling. Finally, the last map corresponds to the cross component and is only related to the quadrupole signal, $\Sigma_{cr} \gamma_\times (x,y)$.}
\label{fig:mapa}
\end{figure*}
The 3D density profile of the neighbouring mass distribution is modelled considering the halo-matter correlation function, $\xi_{hm}$, as:
\begin{equation} \label{eq:rho2h}
    \rho_{2h}(r) = \rho_{m} \xi_{hm} = \rho_{\rm crit} \Omega_m (1+z)^3 b(M_{200},\langle z \rangle) \xi_{mm}
\end{equation}
where $\rho_m$ is the mean density of the Universe ($\rho_m = \rho_{\rm crit} \Omega_m (1+z)^3$) and the halo-matter correlation function is related to the matter-matter correlation function through the halo bias \citep[$\xi_{hm}  = b(M_{200},\langle z \rangle) \xi_{mm}$,][]{Seljak2004}. We set the halo bias by adopting \citet{Tinker2010} model calibration.

Therefore, to model the total surface density profile we consider an elliptical distribution for the main halo component with an elongation $\epsilon_{1h}$ plus the term introduced by the neighbouring distribution also elongated and characterised by the aligned ellipticity component, $\epsilon_{2h}$, as:
\begin{equation} 
\label{eq:smodel}
\begin{split}
    \Sigma(R) = \kappa \Sigma_{cr} = & \Sigma_{1h}(r) + \epsilon_{1h} \Sigma^{\prime}_{1h}(r) \cos(2\theta) + \\
    & \Sigma_{2h}(r) + \epsilon_{2h} \Sigma^{\prime}_{2h}(r) \cos(2\theta) 
\end{split}
\end{equation}
where $\Sigma_{1h}$ corresponds to the projected NFW profile (Eq. \ref{eq:nfw}) and $\Sigma_{2h}$ is the projected density of the neighbouring mass (Eq. \ref{eq:rho2h}). The quadrupoles are related to each monopole component with $\Sigma^{\prime}_{1h} = -r d(\Sigma_{1h}(r))/dr$ and $\Sigma^{\prime}_{2h} = -r d(\Sigma_{2h}(r))/dr$. Finally, $\theta$ is the position angle with respect to the major semi-axis of the halo mass distribution as in Eq. \ref{eq:smodel}. The correspondent semi-axis ratios for each component are obtained as $q_{1h}= (1-\epsilon_{1h})/(1+\epsilon_{1h})$ and $q_{2h}= (1-\epsilon_{2h})/(1+\epsilon_{2h})$.

\subsubsection{Shear components}

The tangential and cross \textit{shear} components related with the observed background galaxy ellipticities, can be obtained from the deflection potential corresponding to the defined mass distribution and can also be decomposed into the monopole and quadrupole contributions:
\begin{align} \label{eq:gamma}
&    \gamma_{\rm{t}} (r,\theta) = \gamma_{\rm{t},0}(r) + \epsilon \gamma_{\rm{t},2}(r) \cos(2\theta),\\
&    \gamma_\times (r,\theta) = \epsilon \gamma_{\times,2}(r) \sin(2\theta). \nonumber
\end{align}

These \textit{shear} components are related with the surface density distribution through \citep{Uitert2017}:
\begin{align} \label{eq:gcomponents}
& \Sigma_{\rm crit} \,  \gamma_{\rm{t},0}(r)  = \frac{2}{r^2} \int^r_0 r^\prime \Sigma_0(r^\prime) dr^\prime - \Sigma_0(r),\\
& \Sigma_{\rm crit} \, \gamma_{\rm{t},2}(r) = -\frac{6 \psi_2(r)}{r^2} - 2\Sigma_0(r) - \Sigma_2(r) \nonumber, \\
& \Sigma_{\rm crit} \, \gamma_{\times,2}(r) = -\frac{6 \psi_2(r)}{r^2} - 4\Sigma_0(r),  \nonumber
\end{align}
where $\psi_2(r)$ is the quadrupole component of the lensing potential and is obtained as:
\begin{equation}\label{eq:psi2}
    \psi_2(r) = -\frac{2}{r^2} \int_0^r r^{\prime 3} \Sigma_0(r^\prime) dr^{\prime}.
\end{equation}
By averaging the tangential \textit{shear} in \ref{eq:gamma} in annular bins we obtain the usually defined density contrast
 $\Delta \Sigma$,
\begin{equation} \label{eq:DSigma}
\Delta \Sigma(r) =  \Sigma_{\rm crit} \gamma_{\rm{t},0}(r) = \frac{1}{2\pi} \int_0^{2\pi} \Sigma_{\rm crit}\, \gamma_{\rm{t}} (r,\theta) d\theta,  
\end{equation}
equivalent to the specified in Eq. \ref{eq:gcomponents}, which is the only term observed in the case of an axis-symmetric mass distribution and it is only related with the monopole.

If we average the $\gamma_t$ and $\gamma_\times$ projections in annular bins, we can isolate the quadrupole components scaled according to the ellipticity:
\begin{align} \label{eq:gproj1}
& \Gamma_{\rm{T}}(r) := \epsilon \Sigma_{\rm crit} \gamma_{\rm{t},2}(r) = \frac{1}{\pi} \int_0^{2\pi} \Sigma_{\rm crit} \gamma_{\rm{t}} (r,\theta) \cos(2\theta) d\theta, \\
\label{eq:gproj2}
& \Gamma_{\times}(r) := \epsilon \Sigma_{\rm crit} \gamma_{\times,2}(r) = \frac{1}{\pi} \int_0^{2\pi} \Sigma_{\rm crit} \gamma_\times (r,\theta) \sin(2\theta) d\theta.
\end{align}
Here we define the distance independent quantities related with the quadrupole, $\Gamma_{\rm{T}}$ and $\Gamma_\times$.
We model the \textit{shear} profiles described in Eq. \ref{eq:DSigma}, \ref{eq:gproj1} and \ref{eq:gproj2}, considering the projected surface density as the sum of the surface components, $\Sigma_0 = \Sigma_{1h} + \Sigma_{2h}$, computed by using \textsc{COLOSSUS}\footnote{\href{https://bitbucket.org/bdiemer/colossus/src/master/}{https://bitbucket.org/bdiemer/colossus/src/master/}} astrophysics toolkit \citep{Diemer2018}. 

\begin{figure*}
    \includegraphics[scale=0.6]{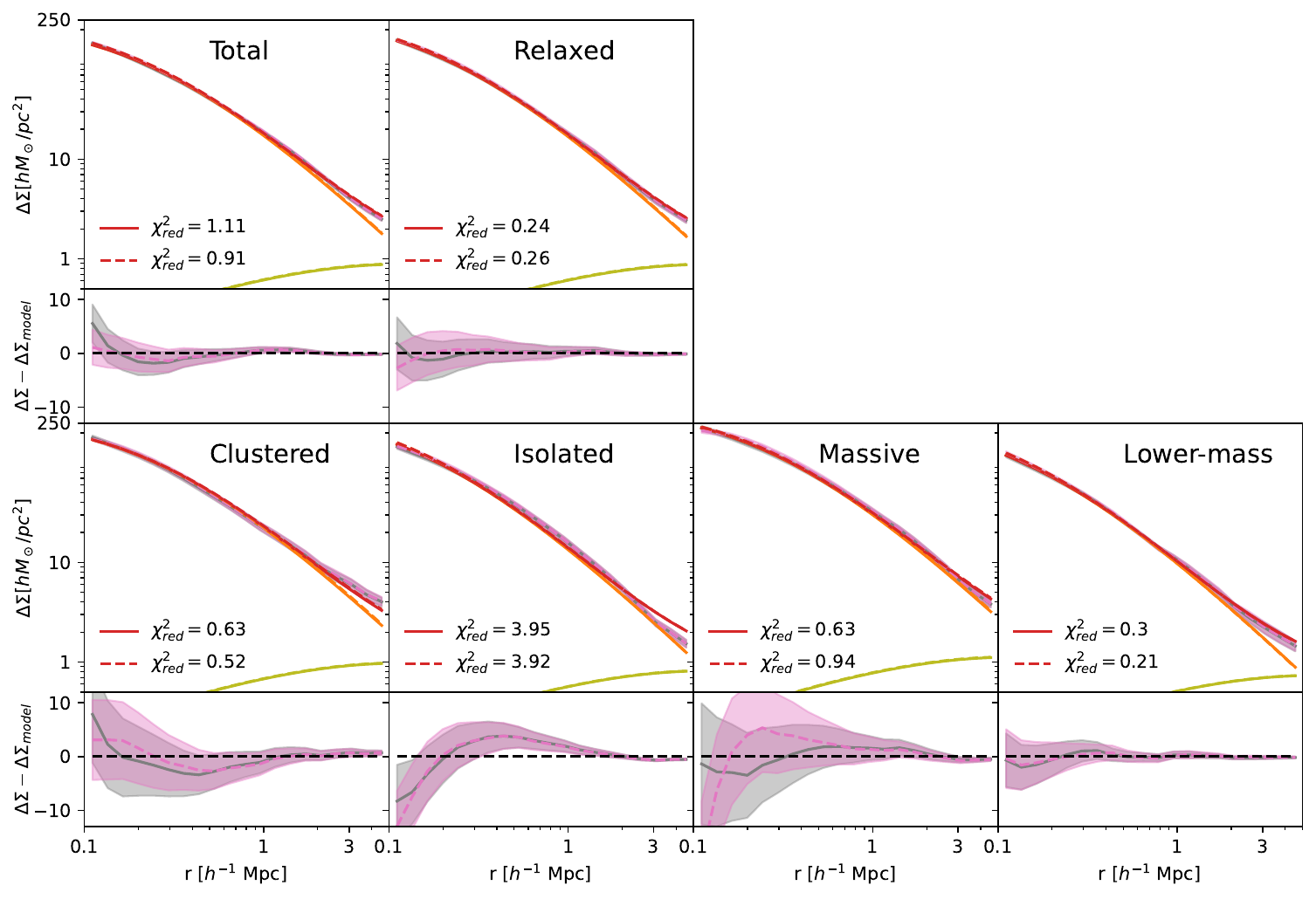}
    \caption{Fitted contrast density profiles, $\Delta \Sigma$, for the halo subsets analysed. In solid grey and dashed pink are the profiles obtained for the CDM and SIDM halos, respectively. Fitted relations are shown in solid and dashed lines, which are almost identical. In orange and light-green we show the corresponding 1-halo and 2-halo components, and the red lines represent the sum of these two components. The panels below each profile shows the differences between the observed contrast density and the fitted model (red solid lines).}
    \label{fig:DSprofiles}
\end{figure*}

\subsection{Profile computation and estimators}

\textit{Shear} profiles are computed from the surface density maps, $\Sigma(x,y)$, of the stacked halos, which are related to the convergence through Eq. \ref{eq:kappa}. We first obtain the stacked $\Sigma(x,y)$ maps, by considering all the particles within a $20 h^{-1} \rm Mpc$ size box, centred in each halo and projected along the three Cartesian axes of the simulation. In this way, we account for the line-of-sight contribution present in the lensing studies and we increase the number of halos considered in the analysis. Before all the particles are combined to compute the maps, we rotate their positions taking into account the main orientation, $\phi$, based on the particle distribution of each halo (see \ref{subsec:hshape}). Therefore, the main direction of the stacked map is aligned with the horizontal axis, $x$. $\Sigma(x,y)$ is then obtained by computing the particle density of all the considered halos within a grid of $8 \times 8 \,(h^{-1}\rm Mpc)^{2}$ area divided in 500$\times$500 bins, leading to a pixel resolution of $\sim 32 \times 32 (h^{-1}\, \rm kpc)^2$. This map is re-scaled by dividing each pixel density by the number of projected halos combined. This procedure is done for all the halo subsets defined in Table \ref{tab:sampdef}. 

Shear maps are then obtained from the density maps using \texttt{lenspac}\footnote{\href{https://github.com/CosmoStat/lenspack}{ https://github.com/CosmoStat/lenspack}} package, which applies the inverse of the KS-inversion \citep{Kaiser1993} to compute the $\gamma$ components. 
For illustration, in Fig. \ref{fig:mapa} we show the stacked maps derived for the total sample of halos analysed in SIDM and CDM simulations. It can be seen that the differences between the maps are more significant in the quadrupole components.

Using these tangential and cross components maps of the \textit{shear}, we compute the lensing estimators by averaging the pixel values included in annular bins:
\begin{align} 
& \widetilde{\Delta \Sigma}(r) = \frac{\sum_{j=1}^{N_\text{pix}} (\Sigma_{{\rm crit}} \gamma_{\rm t})_j}{N_\text{pix}}, \label{eq:profDS} \\
& \widetilde{\Gamma}_{\rm{T}}(r) = \frac{\sum_{j=1}^{N_\text{pix}}  (\Sigma_{{\rm crit}} \gamma_{\rm t})_j \cos{2 \theta_j}}{\sum_{j=1}^{N_\text{pix}}  \cos^2{2\theta_j}}, \label{eq:profgammatan}\\
& \widetilde{\Gamma}_{\times}(r) = \frac{\sum_{j=1}^{N_\text{pix}}  (\Sigma_{{\rm crit}} \gamma_{\rm \times})_j \sin{2\theta}_j}{\sum_{j=1}^{N_\text{pix}} \sin^2{2\theta_j}}.\label{eq:profgammax}
\end{align}
Here the sums runs over the $N_\text{pix}$ pixels which centres are included within a radial bin $r\pm\delta r$.  $(\Sigma_{{\rm crit}} \gamma_{\rm t})_j$ and $(\Sigma_{{\rm crit}} \gamma_{\times})_j$ are the $j-$pixel value of tangential and cross \textit{shear} maps, respectively, and $\theta_j$ is the position angle of the $j-$pixel with respect to the $x-$axis. Profiles are computed considering 20 logarithmic annular bins from $100h^{-1}$\,kpc up to $5 h^{-1}$\,Mpc. The respective errors for each profile, $\sigma_{\Delta \Sigma}$, $\sigma_{\Gamma_\text{T}}$ and $\sigma_{\Gamma_\times},$ are computed using a bootstrap resampling with 50 iterations over the total number of halos in each subset. In order to compute the errors we produce for each halo sample 50 resampled maps.

\begin{figure*}
    \includegraphics[scale=0.5]{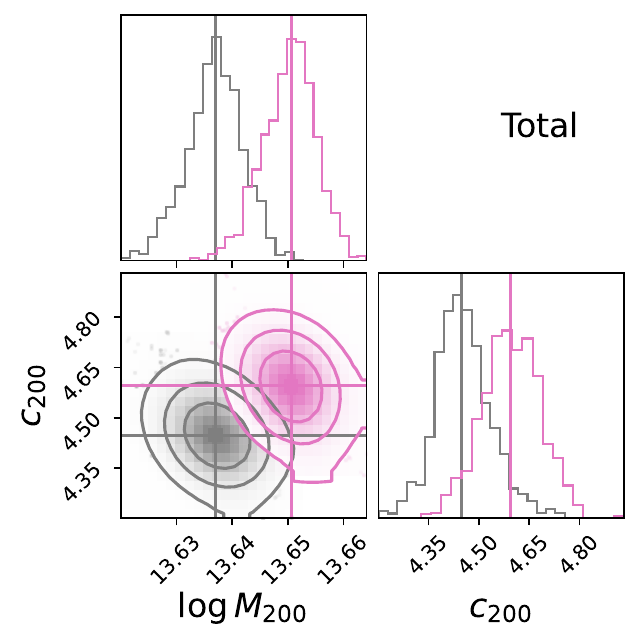}
    \includegraphics[scale=0.5]{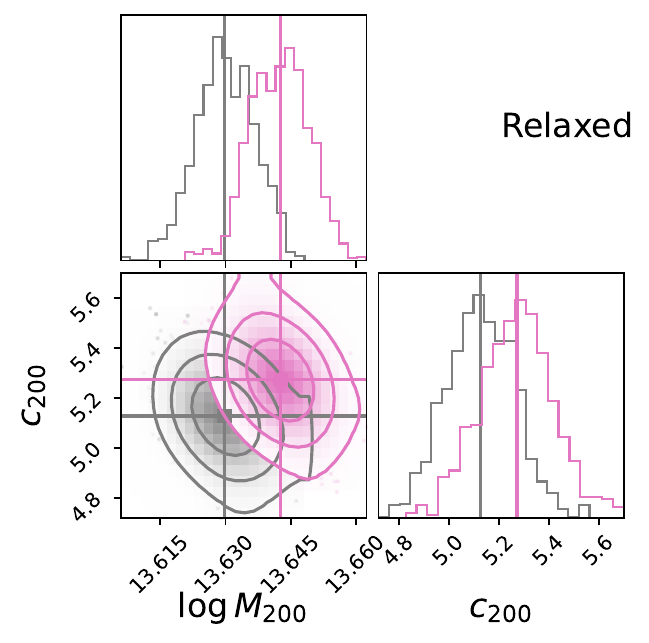}
    \includegraphics[scale=0.5]{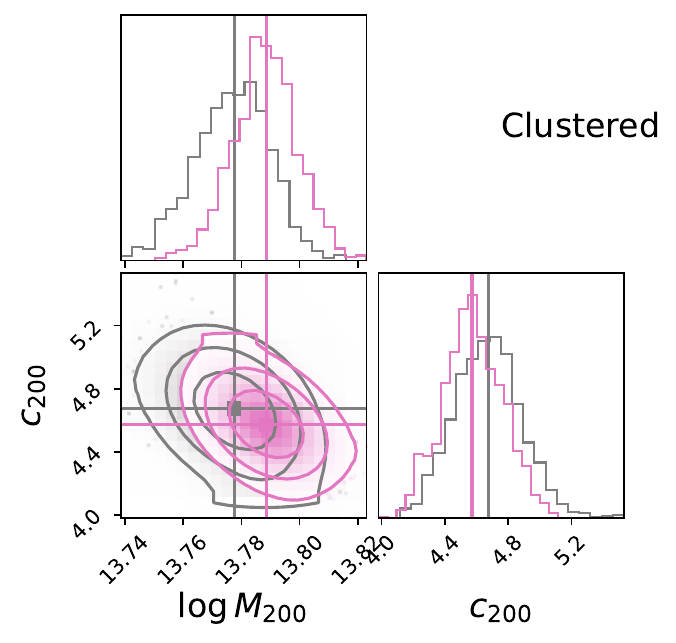}
    \includegraphics[scale=0.5]{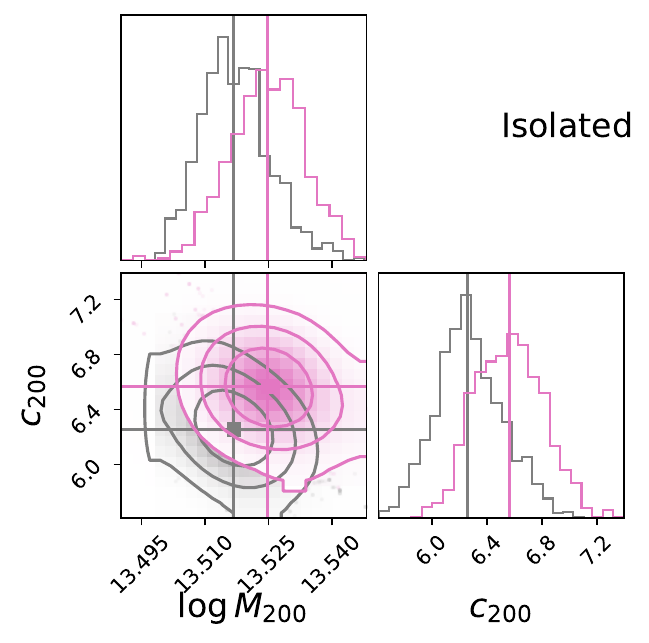}
    \includegraphics[scale=0.5]{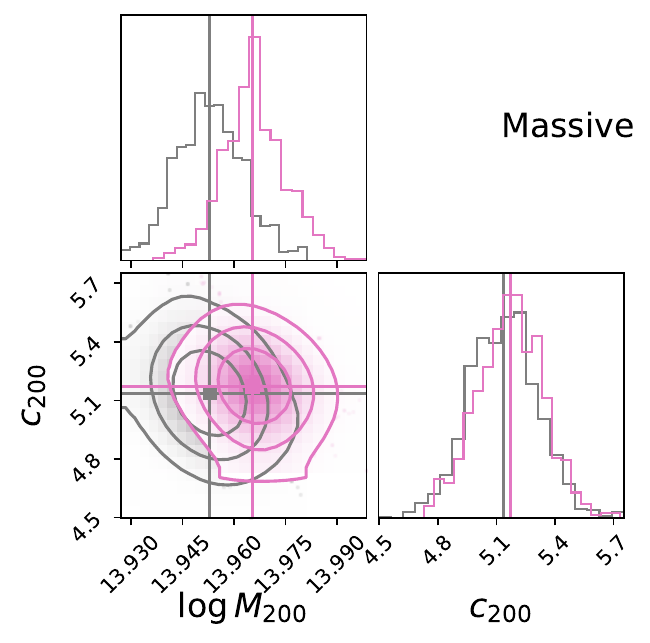}
    \includegraphics[scale=0.5]{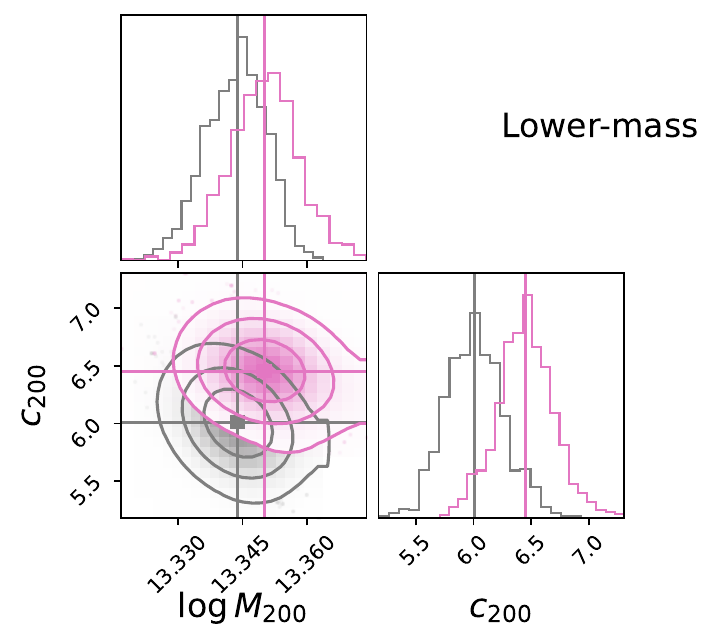}
    \caption{Posterior density distributions for the fitted parameters from the contrast density profiles, $\Delta \Sigma$, mass and concentration. In grey and pink are the results for the CDM and SIDM, respectively. The distributions are shown after discarding the first 50 steps of each chain. Vertical lines indicate the median values of the distributions adopted as $\log M_{200}$ and $c_{200}$.}
    \label{fig:mcmc_mono}
\end{figure*}

\section{Results}
\label{sec:results}

We perform the fitting procedure in order to obtain the parameters that characterise the surface density distribution, in the same way as it would be conducted using the observational data. First, we obtain the mass and the concentration parameters by fitting the contrast density profiles, $\Delta \Sigma$, since this profile is sensitive only to the monopole component of the modelling and has a larger signal-to-noise ratio compared with the quadrupole profiles. After that, we constrain the projected halo shapes by fitting the quadrupole components. 
In this section, we begin by presenting the results related to the monopole component and then we present the halo shapes derived according to the quadrupole components, by considering two different models.  

\subsection{Fitted parameters from the monopole}

 We constrain fit the contrast density profiles by using the Markov chain Monte Carlo (MCMC) method, implemented through \texttt{emcee} python package \citep{Foreman2013}, to optimise the  log-likelihood function for the monopole profile:
\begin{equation}
\label{eq:loglmono}
\ln{\mathcal{L}}(\Delta \Sigma | M_{200},c_{200}) = \frac{1}{2} \frac{(\widetilde{\Delta \Sigma} - \Delta \Sigma)^2}{\sigma^2_{\Delta \Sigma}} + \ln{2 \pi \sigma^2_{\Delta \Sigma}},
\end{equation}
where $\widetilde{\Delta \Sigma}$ is the profile computed from the tangential shear map according to Eq. \ref{eq:profDS}, $\sigma_{\Delta \Sigma}$ its respective bootstrap error and $\Delta \Sigma$ is the model (Eq. \ref{eq:DSigma}) computed considering $\Sigma_0$ as the sum of the main and second halo components,  $\Sigma_0 = \Sigma_{1h} + \Sigma_{2h}$. To fit the data we use 15 steps and 250 steps with flat priors,  $12.5 < \log(M_{200}/(h^{-1} M_\odot)) < 15.5$, $4 < c_{200} < 8$. Our best fit parameters are obtained after discarding the first 50 steps of each chain, according to the median of the marginalised posterior distributions and errors enclose the central $64$\% of the marginalised posterior.

Fitted profiles are shown in Fig. \ref{fig:DSprofiles} together with the reduced chi-squares values. The adopted modelling is suitable for all the sub-sets but for the Isolated sample of halos, in which the amplitude of the 2-halo term overestimates the surface density distribution at larger radial scales. Fitted posterior density distributions for each sample are shown in Fig. \ref{fig:mcmc_mono}. Mass and concentrations derived for each combined halo sample are in general in agreement within the errors for both simulated data-sets. However, profiles from the SIDM simulation tend to describe more massive (by $\sim 1\%$) and concentrated (by $\sim 3\%$) distributions. Previous studies have shown that density profiles in SIDM simulations tend to show higher concentrations which can be related with a mass displacement from the inside out, enhancing the density of SIDM halos around or beyond $r_s$, resulting in a steeper profile at such radii \citep{Banerjee2020}, thus in agreement with our findings. Although with low significance, differences in the masses might be related with the well-known interplay between mass and concentration \citep[e.g. ][]{Bullock2001,Duffy2008,Bhattacharya2013,Okoli2017,Ishiyama2021}. The combined relation between these parameters can be also noticed from the contour distributions.

\begin{figure*}
    \includegraphics[scale=0.55]{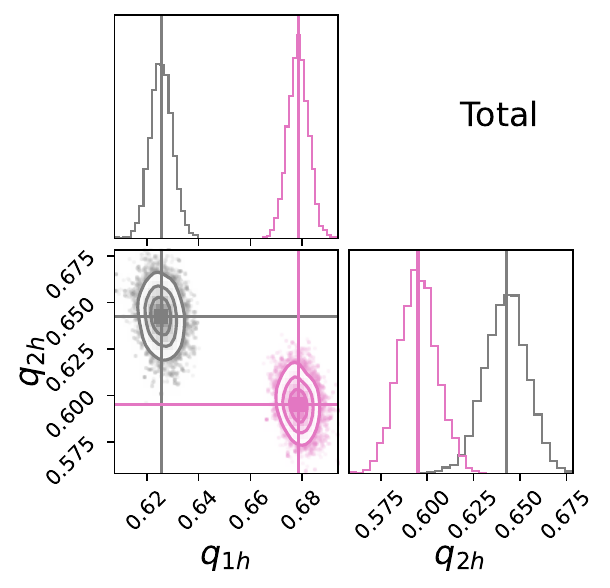}
    \includegraphics[scale=0.55]{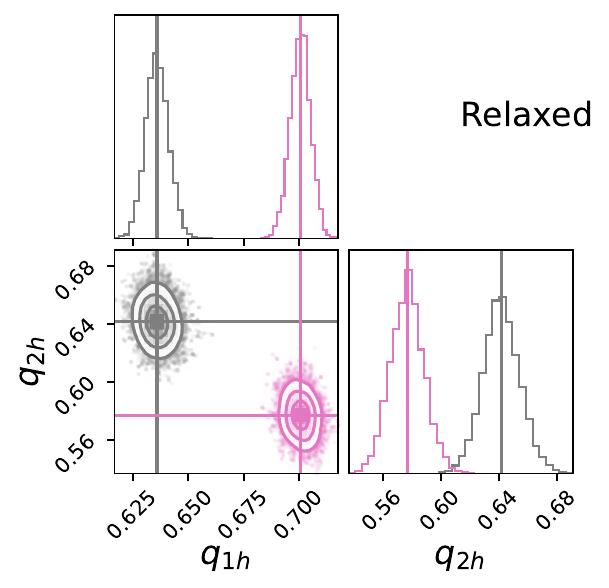}
    \includegraphics[scale=0.55]{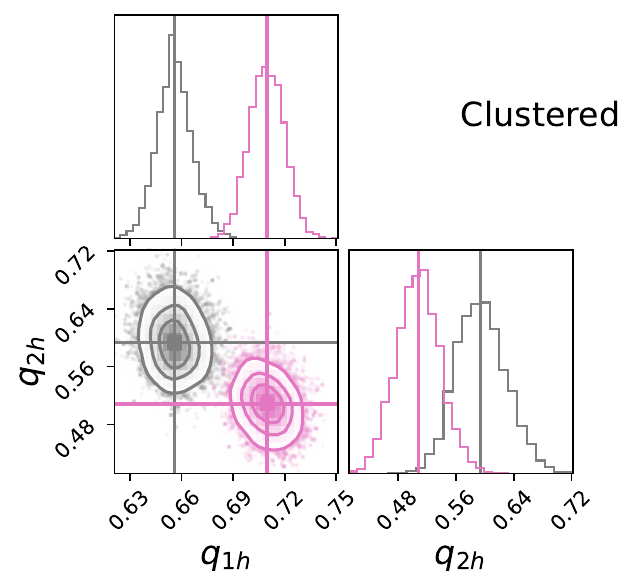}
    \includegraphics[scale=0.55]{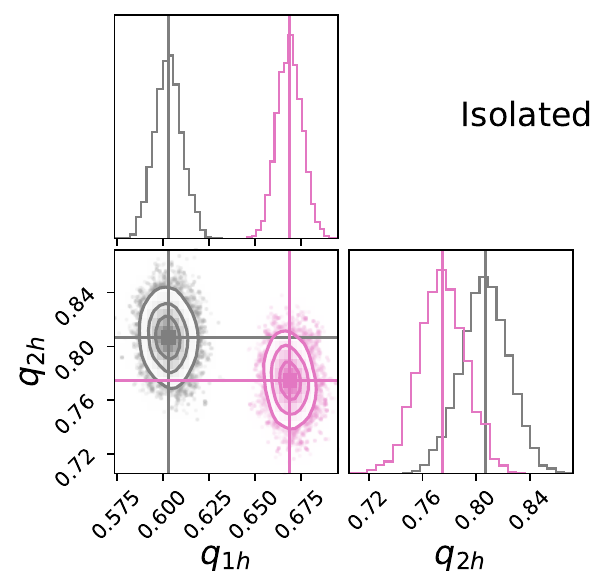}
    \includegraphics[scale=0.55]{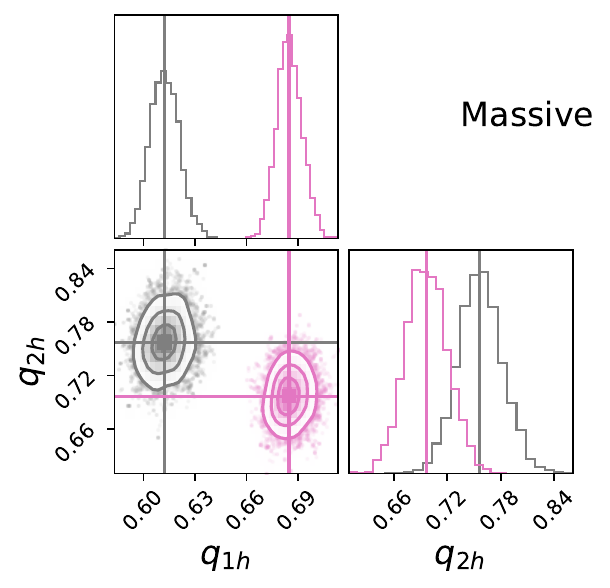}
    \includegraphics[scale=0.55]{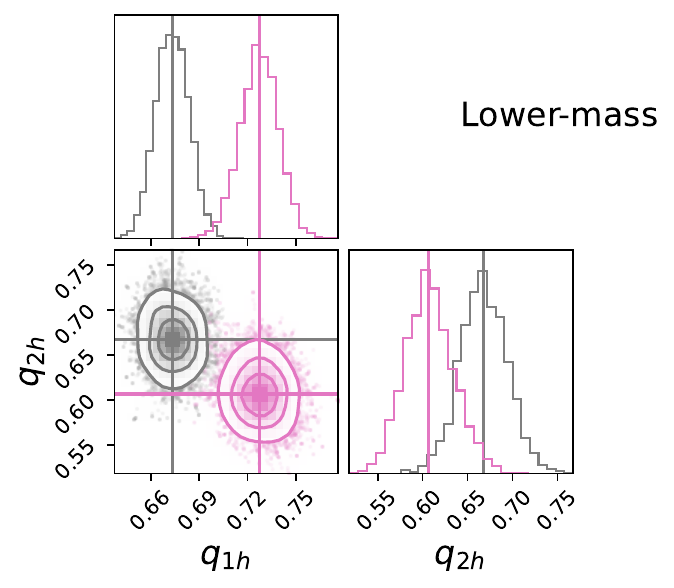}
    \caption{Posterior density distributions for the fitted parameters ($q_{1h}$ and $q_{2h}$) from the tangential and cross quadrupole component profiles, $\Gamma_T$ and $\Gamma_\times$. In grey and pink are the results for the CDM and SIDM, respectively. The distributions are shown after discarding the first 200 steps of each chain and the vertical lines indicate the median values.}
    \label{fig:mcmc_q}
\end{figure*}

\begin{figure*}
    \includegraphics[scale=0.6]{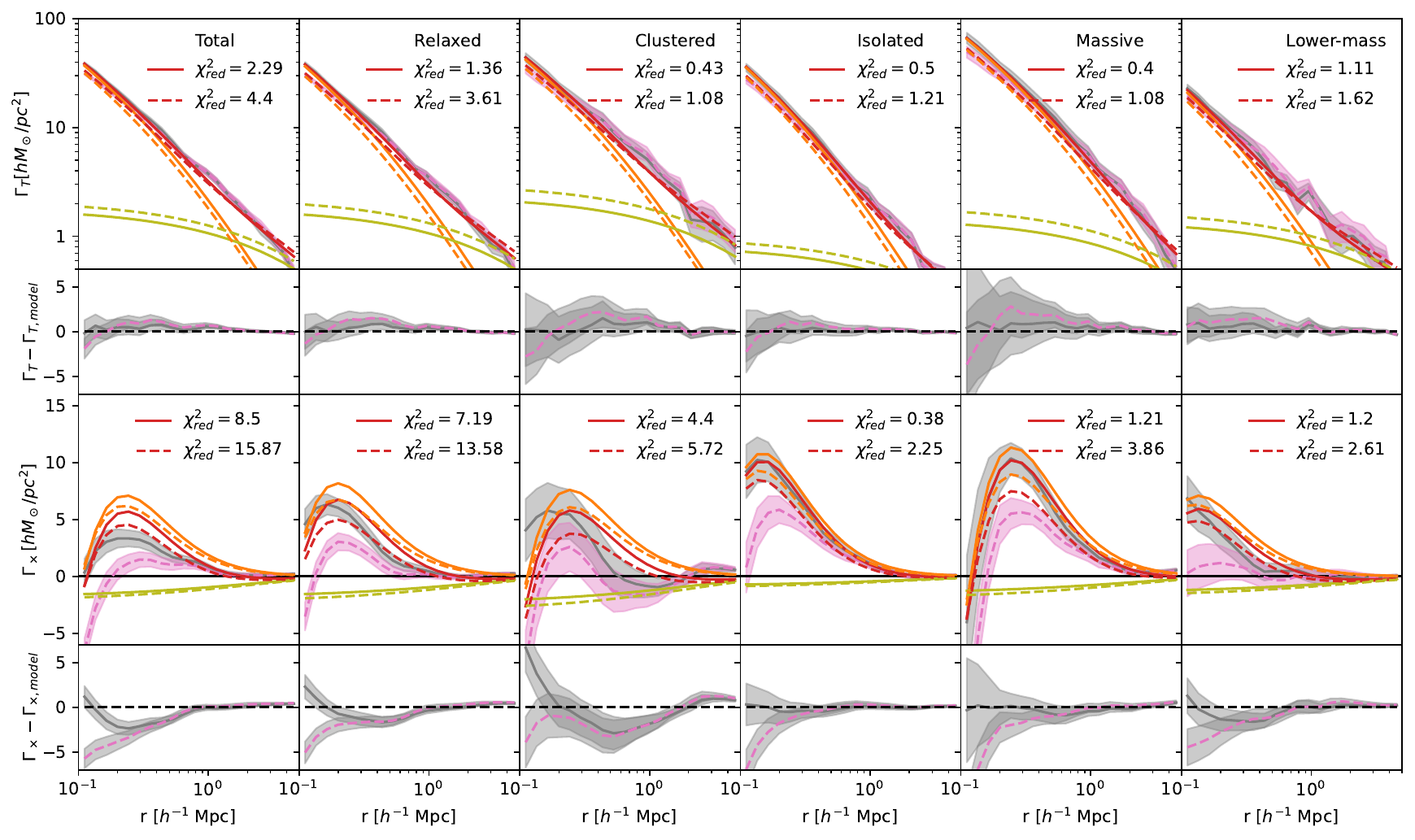}
    \caption{Fitted quadrupole profiles, tangential (upper panels) and cross (lower panels) components, for the halo subsets analysed. In solid grey and dashed pink are the profiles obtained for the CDM and SIDM halos, respectively. The shadow region express the obtained errors according to bootstrap resampling. Fitted relations are shown in solid and dashed lines. In orange and light-green we show the corresponding to the 1-halo and 2-halo components, respectively, and the red lines represent the sum of these two components. $q_{1h}$ and $q_{2h}$ are obtained according to the fitting procedure described in \ref{subsec:fitted}. The panels below each profile shows the differences between the observed quadropole profiles and the fitted model (red solid lines). }
    \label{fig:Qprofiles}
\end{figure*}

\begin{figure}
    \includegraphics[scale=0.6]{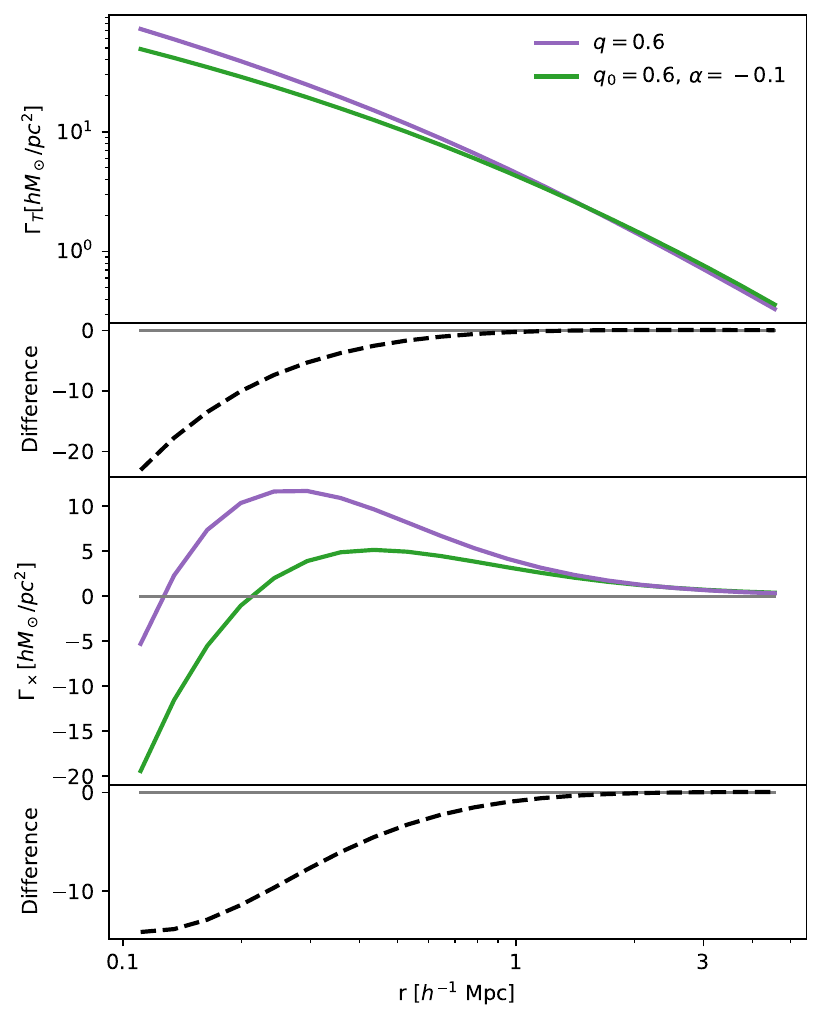}
    \caption{Quadupole profiles for a NFW surface density distribution with $M_{200} = 10^{14} M_\odot$ and $c_{200} = 5$ elongated by a fix semi-axis ratio, $q = 0.6$ (purple solid line) and by a radial dependent semi-axis ratio $q = q_0 r^\alpha$ (green solid line). Bottom panels show the difference between both profiles (black dashed line). }
    \label{fig:radial}
\end{figure}

\subsection{Mean aligned elongation constrained from the quadrupoles}
\label{subsec:fitted}
Once we have constrained $M_{200}$ and $c_{200}$ for each halo subset, we proceed to simultaneously fit the quadrupole components using a similar approach, minimising the sum of the likelihoods $\ln{\mathcal{L}}(\Gamma_{\rm{T}} | r ,\tilde{q}_{1h}, \tilde{q}_{2h}) + \ln{\mathcal{L}}(\Gamma_{\times} | r ,\tilde{q}_{1h}, \tilde{q}_{2h})$, defined as:
\begin{equation} \label{eq:loglqt}
\ln{\mathcal{L}}(\Gamma_{\rm{T}} | \tilde{q}_{1h}, \tilde{q}_{2h}) = \frac{1}{2} \frac{( \widetilde{\Gamma}_{\rm{T}} - \Gamma_{\rm{T}})^2}{\sigma_{\Gamma_\text{T}}} + \ln{2 \pi \sigma^2_{\Gamma_\text{T}}} 
\end{equation}
\begin{equation} \label{eq:loglqx}
\ln{\mathcal{L}}(\Gamma_{\times} | \tilde{q}_{1h}, \tilde{q}_{2h}) = \frac{1}{2} \frac{( \widetilde{\Gamma}_\times - \Gamma_\times)^2}{\sigma_{\Gamma_\times}} + \ln{2 \pi \sigma^2_{\Gamma_\times}}, 
\end{equation}
where $\widetilde{\Gamma}_{\rm{T}}$ and $\widetilde{\Gamma}_\times$ are the profiles computed from the tangential and cross \textit{shear} maps according to Equations \ref{eq:profgammatan} and \ref{eq:profgammax}, respectively, $\sigma_{\Gamma_\text{T}}$ and $\sigma_{\Gamma_\times}$ their respective bootstrap error. $\Gamma_{\rm{T}}$ and $\Gamma_{\times}$ are the adopted models (Eq. \ref{eq:gproj1} and Eq. \ref{eq:gproj2}) computed considering  $\Sigma_0 = \Sigma_{1h} + \Sigma_{2h}$. To fit the data we use 15 chains for each parameter and 1000 steps in this case, considering flat priors: $0.6 < \tilde{q}_{1h} < 0.9$ and $0.1 < \tilde{q}_{2h} < 0.5$. The parameters are obtained after discarding the first 200 steps of each chain and the obtained posterior distributions for both fitted parameters are shown in Fig. \ref{fig:mcmc_q}. The fitted semi-axis ratios are tightly constrained and, as expected, they indicate rounder shapes for the halos in the SIDM simulation, with semi-axis ratios roughly $\sim 10 \%$ larger than their counterparts in the CDM simulation. An unexpected result is that the elongation of the neighbouring distribution of the halos tends to be larger for the SIDM simulation. Although the observed differences are not as significant as for the halo elongation, given that $q_{2h}$ has larger errors, halo subsets from the SIDM simulation systematically show larger values,  roughly around $\sim 10 \%$. This result, together with the larger differences obtained for $q_{1h}$ when compared to those based on the particle distribution, might reflect the deficiencies of the model in properly distinguishing between the shapes of both components.
This is also accounted by the large differences between the modelling and the computed profiles quantified from reduced chi-squared values that can be observed in Fig. \ref{fig:Qprofiles}, especially for the cross component and the halos included in the SIDM simulation. 

A potential shortcoming in the adopted modelling approach might be linked to the parameterization used for fitting the main halo component. The impact of the adopted modelling for fitting the quadrupole was already tested in a previous work \citep{Gonzalez2022}, in which we considered three alternative approaches: (1) fitting an Einasto model \citep{Einasto1989,Retana2012} instead of the NFW, (2) fixing the NFW concentration in the analysis by using a concentration relation with mass and redshift and (3) restricting the radial range in the fitting procedure to avoid the impact of the neighbouring component. In that work, we showed that the fitted semi-axis ratios are in general agreement, regardless of the adopted modelling specially when discarding highly un-relaxed halos. This leads us to the conclusion that the constrained shapes are not expected to be significantly influenced by the particular model adopted.
In this work, we extend this analysis by considering an NFW model with a core, since it is expected to better constrain the density distributions observed in SIDM halos. We adopt the model presented in \citep{Neto2007} and fit three parameters to the contrast density profiles, mass, concentration and a parameter that accounts for the presence of a core \citep[$1/b$ in Eq. 2 in][]{Neto2007}. We obtain that the fitted parameters are highly correlated. This is expected since, as was already shown in the previous subsection, there is a correlation between the mass and the concentration that cannot be properly disentangled since the adopted radial range is not highly sensitive to the inner mass distribution. Despite this, we test the impact on the quadrupole fitting by using this modelling. We obtain equivalent results as for the NFW model without a core. The fitted semi-axis ratios are in agreement within $1\%$. The results obtained when varying the model further reinforces  our findings, concluding that the chosen model for the main halo component has a limited impact on the estimated halo shapes based on the quadrupole fitting. However, it is worth noting that in contrast, the constrained masses and concentrations can be significantly influenced by the particular modelling. This result is expected given the low values obtained for the derived $\chi-$squares when fitting $\Delta \Sigma$, indicating that the adopted modelling accurately describes the observed density distributions with highly correlated parameters.

\subsection{Radial variation modelling}
\label{subsec:radial_var}
According to the results presented in \ref{subsec:stacked_shapes}, the shapes of the halo particle distributions show a radial variation, which is more pronounced in the SIDM simulation, as indicated by a steeper fitted power law slope. Taking this into account and the results presented in the previous subsection, we aim to propose a model that accurately captures the shape radial variation. In order to do that, we generate a surface mass map that follows a projected NFW elliptical distribution, $\Sigma (R(q))$, with a radial variation of the semi-axis ratio, $q:=q(r)$, set following Eq. \ref{eq:qr}. From this map, we can compute the expected quadrupole profiles using Equations \ref{eq:profgammatan} and \ref{eq:profgammax}. In Fig. \ref{fig:radial}, we show a comparison between the obtained profiles with and without considering a radial variation for the elongation of the density distribution. As it can be noticed, a radial variation of the semi-axis ratio can significantly affect the quadrupoles, especially at lower radial ranges, where the signal is suppressed due to the rounder distribution towards the centre. 

In this case, we minimise the sum of the likelihoods defined in Equations \ref{eq:loglqt} and \ref{eq:loglqx}, however, the models are modified. For the main halo component, the model for the quadrupole profiles is derived from surface density maps of an elliptical NFW model with a radial variation, $\Sigma [R(q)]$. In order to build the maps, we fix the previously obtained mass and concentrations for each sample, according to the fitted contrast density distributions, $\Delta \Sigma$. We add to this model for the $1-$halo term, the contribution from the neighbouring mass distribution considering the same model as in the previous subsection, i.e. using Eq. \ref{eq:gproj1} and Eq. \ref{eq:gproj2} with  $\Sigma_0 = \Sigma_{2h}$. With this approach, we proceed to fit the quadrupoles considering three free parameters with uniform priors:  $-0.15 < \alpha < 0.15$, $0.4 < q_0 < 0.8$ and $0.3 < q_{2h} < 0.7$. 

\begin{figure*}
    \includegraphics[scale=0.52]{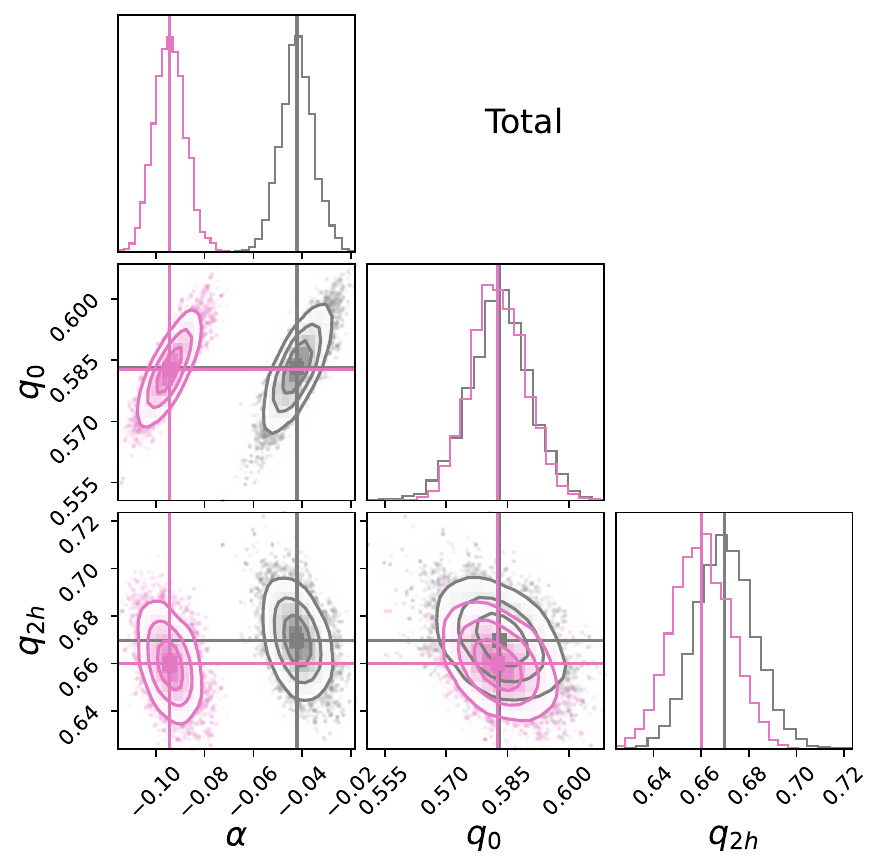}
    \includegraphics[scale=0.52]{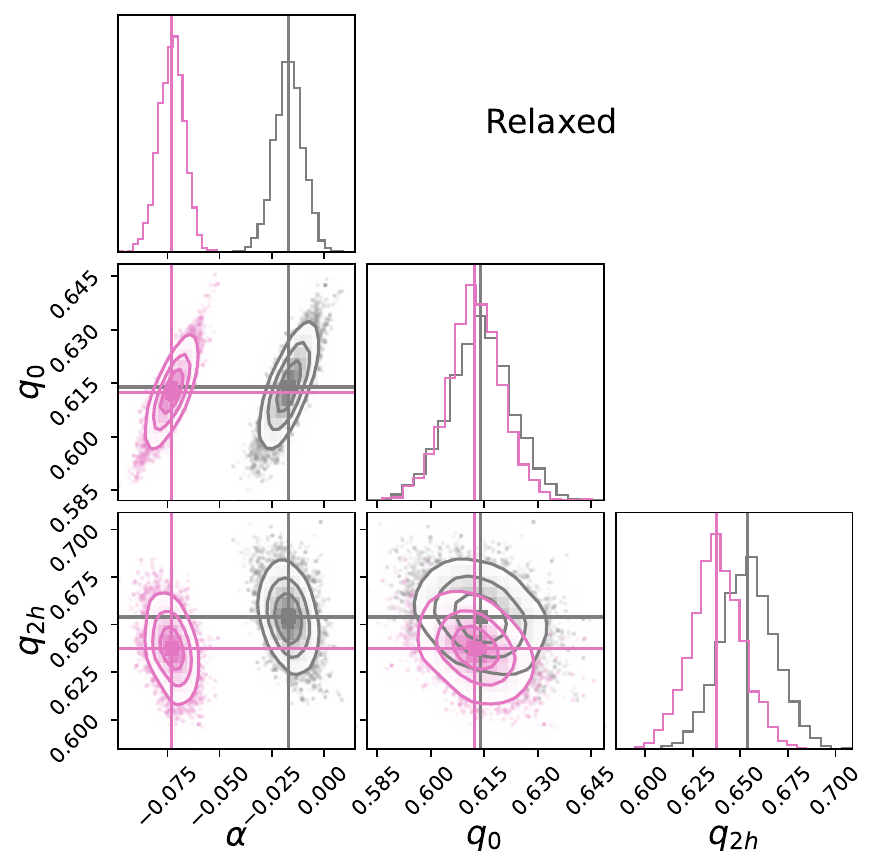}
    \includegraphics[scale=0.52]{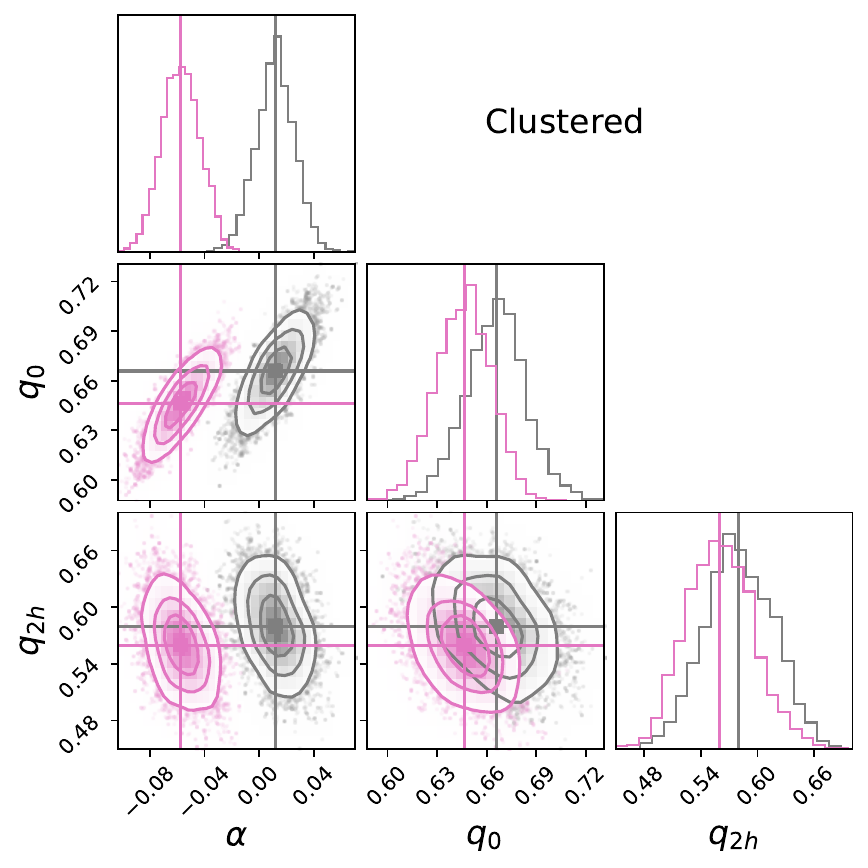}
    \includegraphics[scale=0.52]{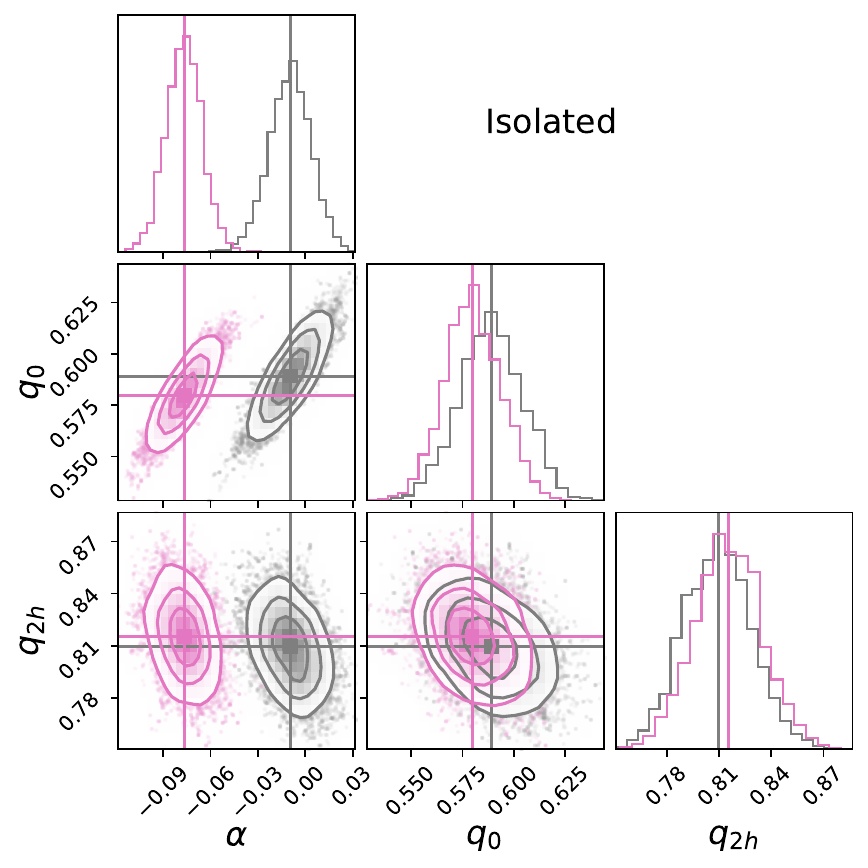}
    \includegraphics[scale=0.52]{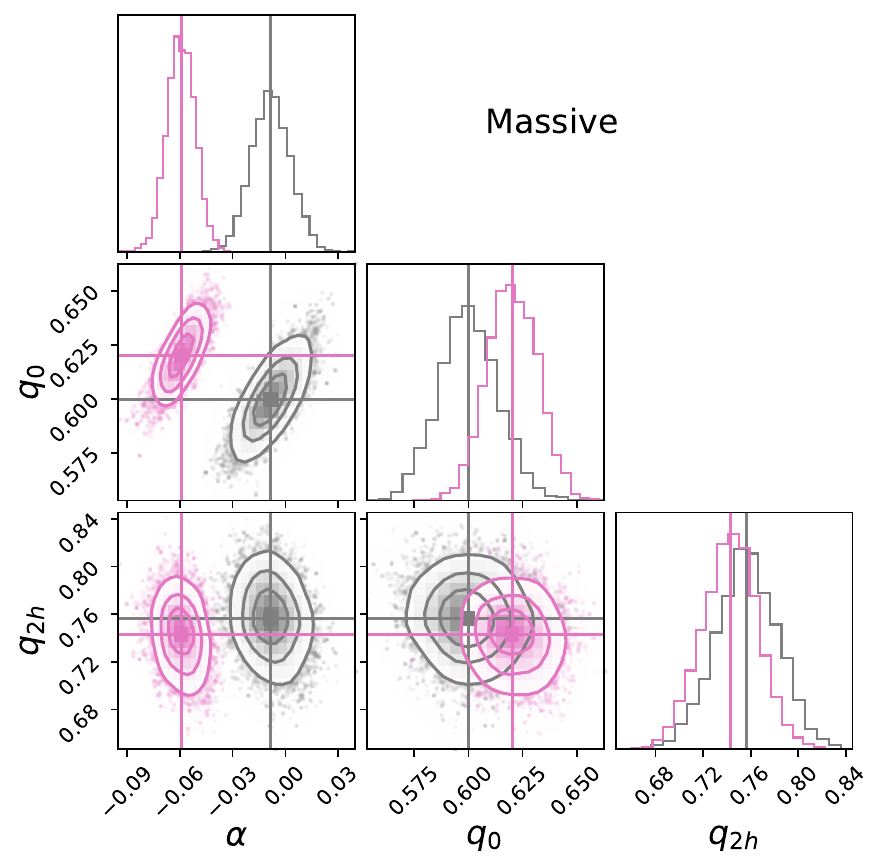}
    \includegraphics[scale=0.52]{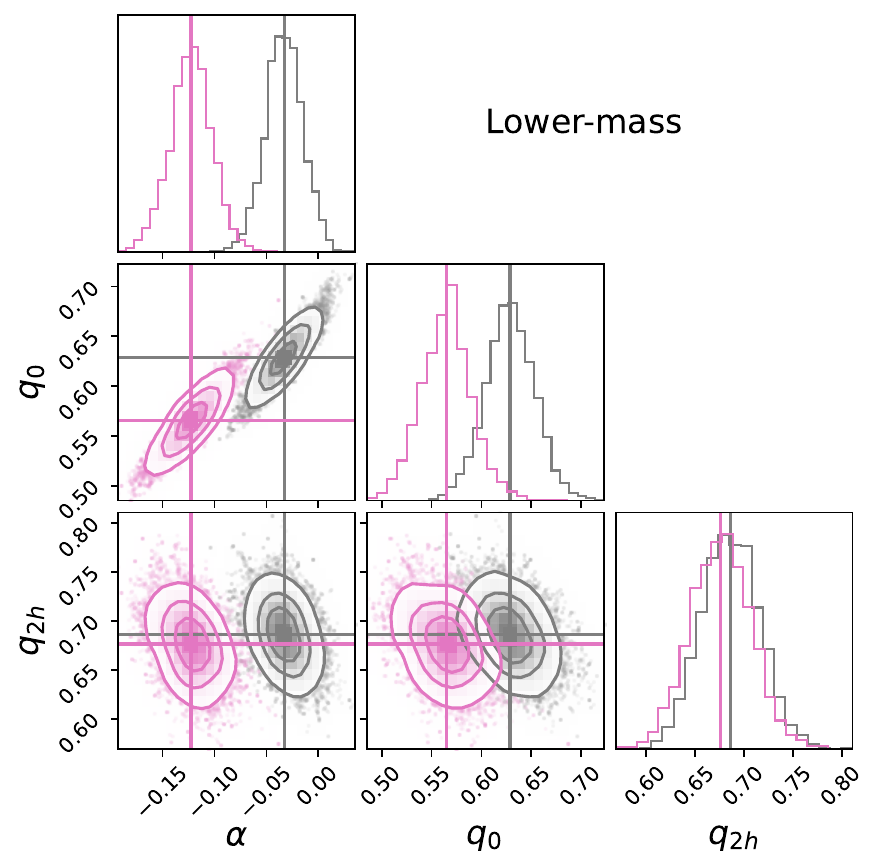}
    \caption{Posterior density distributions for the fitted parameters ($\alpha$, $q_{0}$ and $q_{2h}$) from the tangential and cross quadrupole component profiles, $\Gamma_T$ and $\Gamma_\times$. In grey and pink are the results for the CDM and SIDM, respectively. The distributions are shown after discarding the first 200 steps of each chain and the vertical lines indicate the median values.}
    \label{fig:mcmc_qr}
\end{figure*}

The posterior distributions are presented in Fig. \ref{fig:mcmc_qr}. In this case, it can be noticed that the semi-axis ratios for both simulations are consistent for all the halo subsets analysed. This is evidenced in both, the main halo component, $q_0$, and the neighbouring mass distribution, $q_{2h}$. On the other hand, significant differences are obtained for the power law slope that characterises the radial variation of the elongation. In particular, we obtain a detectable radial variation for the halos in the SIDM simulation, while the halos in the CDM simulation exhibit a negligible variation, compatible with $\alpha = 0$. The proposed model seems to improve the fitting of the cross-quadrupole component especially for the halos in the SIDM simulation, resulting in lower chi-square values. However, according to the contours of the posterior distributions, there is also a noticeable interplay between the fitted parameters $q_0$ and $\alpha$ which reflects that the signal is sensitive to a combination of these parameters. A deeper discussion of these results is presented in the next section. 

\begin{figure*}
    \includegraphics[scale=0.6]{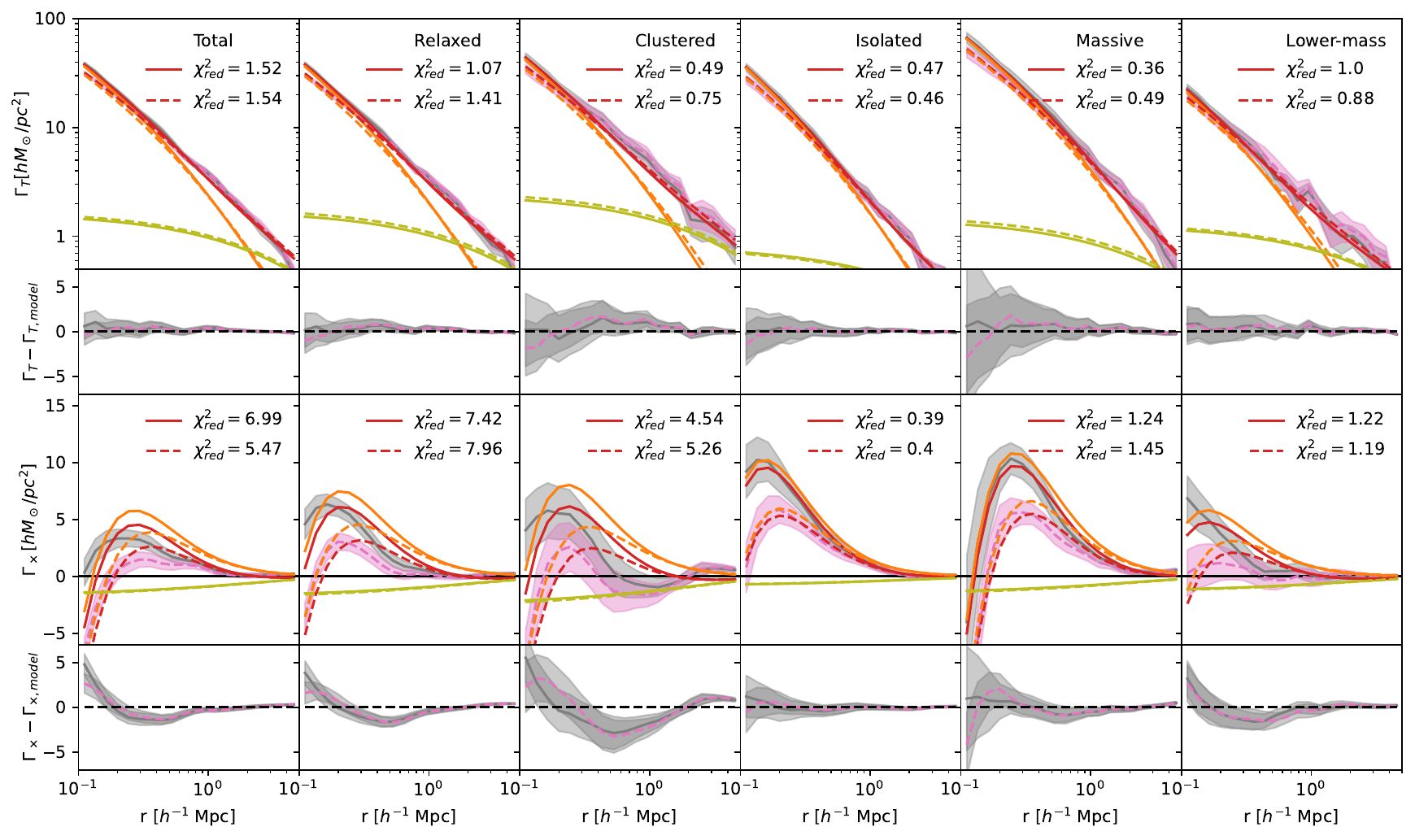}
    \caption{Same as in Fig. \ref{fig:Qprofiles}. $q_{0}$, $\alpha$ and $q_{2h}$ are obtained according to the fitting methodology described in (\ref{subsec:radial_var}).}
    \label{fig:Qprofilesr}
\end{figure*}

\begin{figure*}
    \includegraphics[scale=0.6]{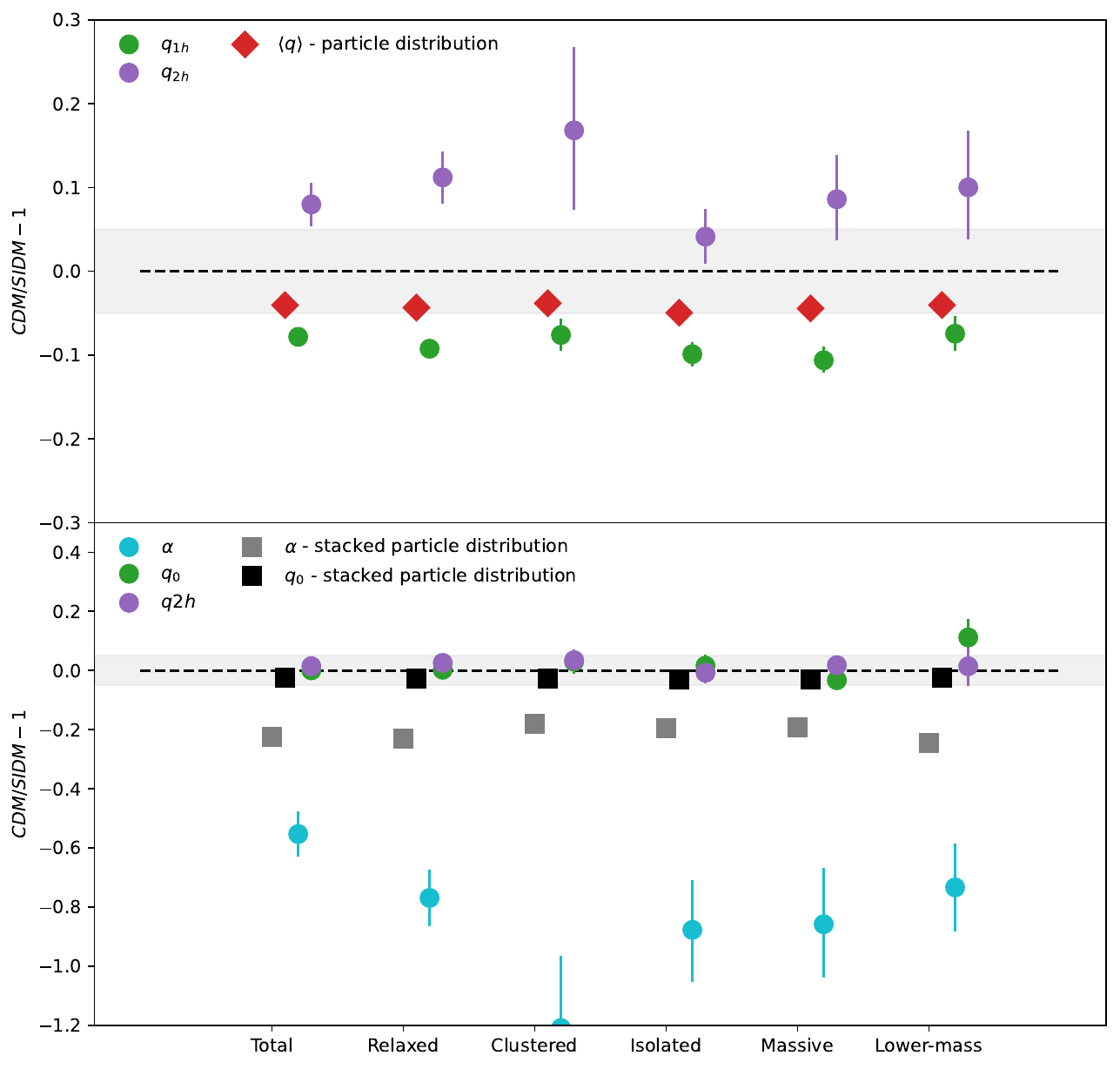}
    \caption{Ratio between the fitted shapes parameters from each sample of halos tacked from the CDM and SIDM simulations. The shadow grey region corresponds to a $5\%$ agreement. \textit{Upper panel:} Red diamonds represent the mean semi-axis ratio obtained for the stacked halos, according to the bound particle distribution using the iterative method (\ref{subsec:hshape}). Green and purple dots correspond to the fitted $q_{1h}$ and $q_{2h}$ from the quadrupole components (\ref{subsec:fitted}) \textit{Bottom panel:} Black and grey squares correspond to $\alpha$ and $q_0$ fitted according to the radial variation of the stacked bound particle distribution (\ref{subsec:stacked_shapes}). Light-blue, green and purple correspond to the fitted parameters, $\alpha$, $q_0$ and $q_{2h}$, according to the quadrupole profiles (\ref{subsec:radial_var}).}
    \label{fig:comparison}
\end{figure*}

\begin{figure*}
    \includegraphics[scale=0.6]{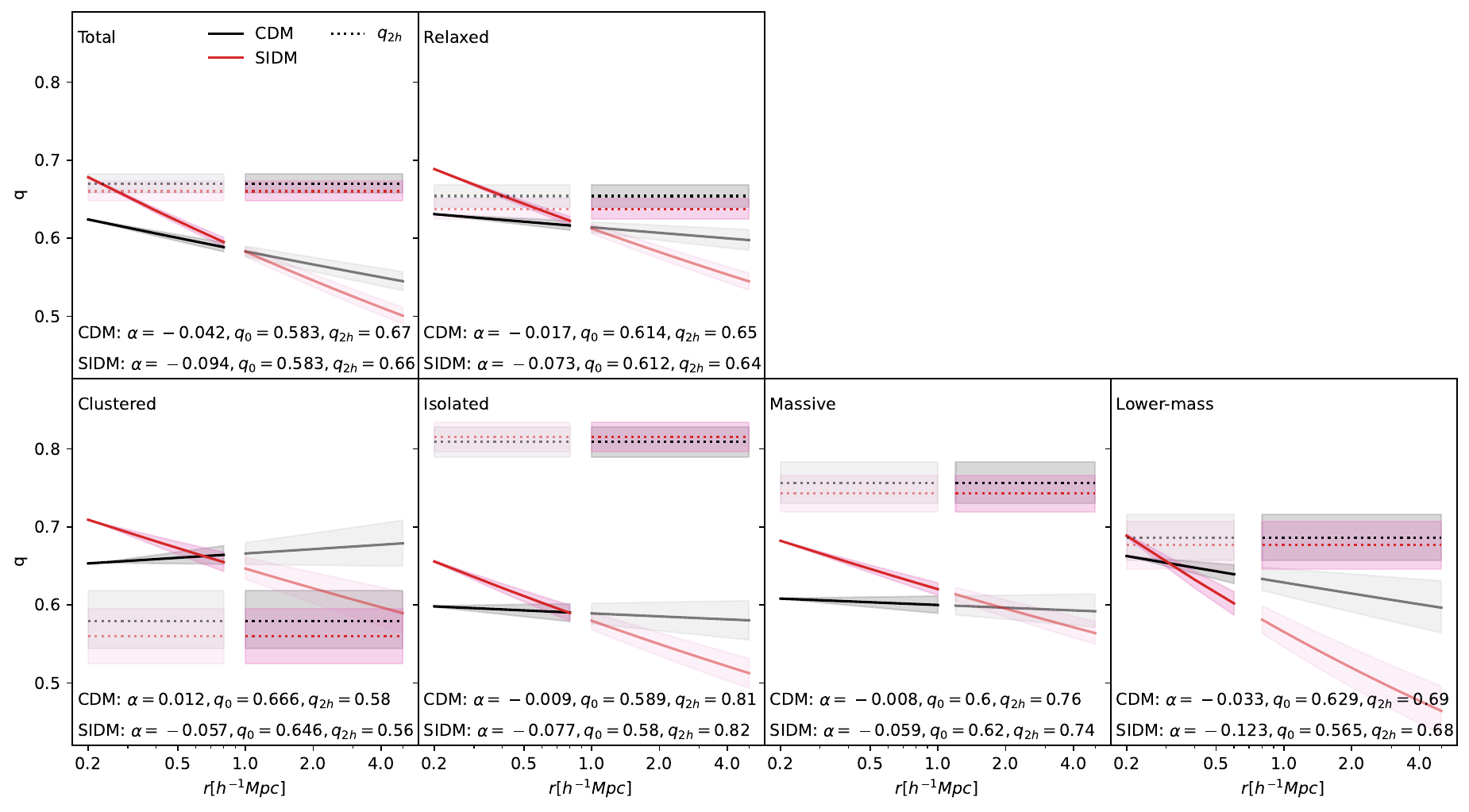}
    \caption{Radial variation of the semi-axis ratio predicted according to the fitted quadrupole components for the main halo (solid lines) and for the neighbouring mass distribution (dotted lines). Results are shown for the stacked halos in the SIDM and CDM simulations (pink and black, respectively). Shadow regions correspond to the expected sampling error computed using the bootstrapping technique. Darker colours indicate the radial ranges in which the components are expected to be more dominant and are set at the mean $R_\text{vir}$ of the stacked halos.}
    \label{fig:result}
\end{figure*}

\begin{figure*}
    \includegraphics[scale=0.6]{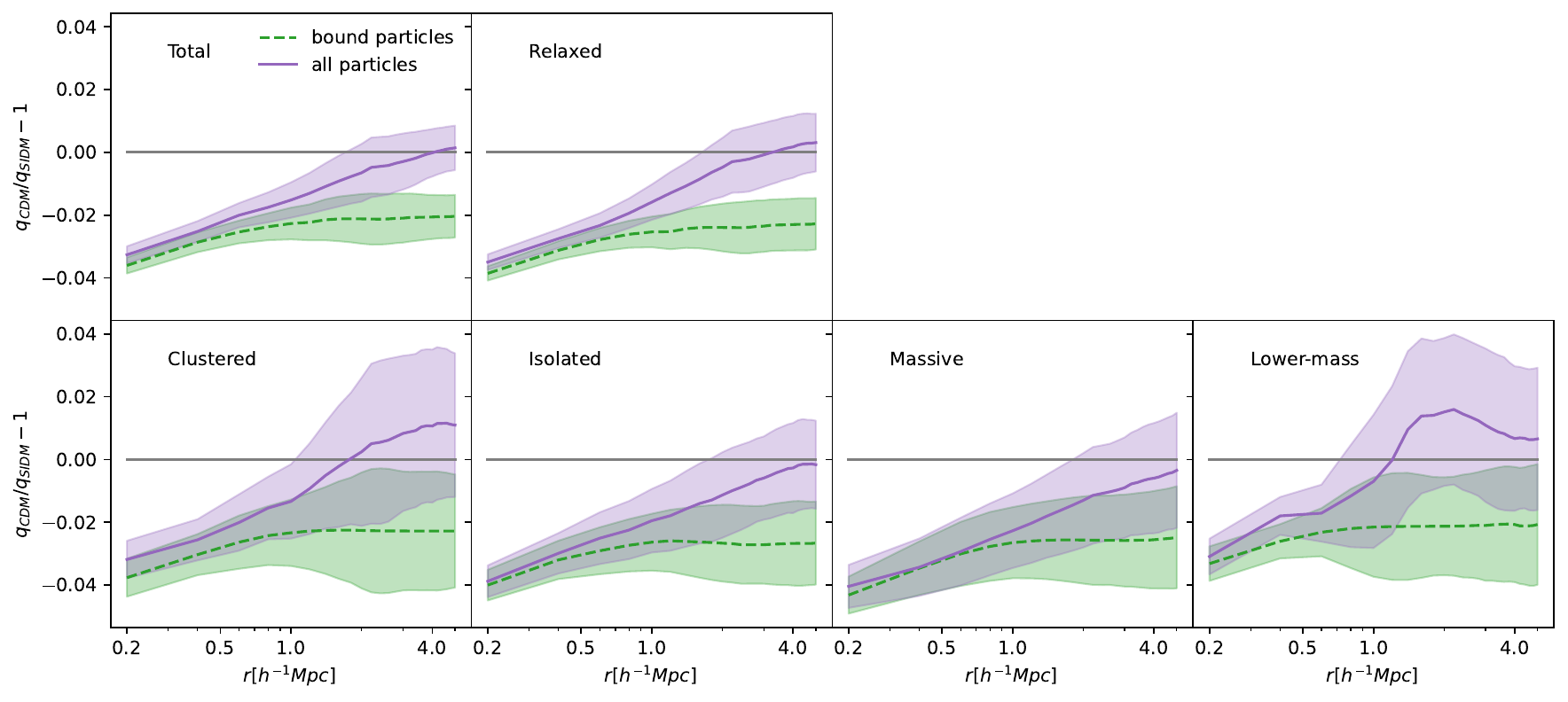}
    \caption{Ratio between the fitted halo semi-axis ratio from the stacked DM particle distribution enclosed within the projected radius $r$. We show in green and dashed lines the results from stacking only the bound particles of the halos and in purple solid line the correspondent for the whole particle distribution within a 10 Mpc$ h^{-1}$ box centred in each halo. Shadow regions correspond to the expected sampling error computed using the bootstrap resampling technique.}
    \label{fig:ratio_DM}
\end{figure*}

\section{Discussion}
\label{sec:discussion}

Allowing strong elastic self-interactions of the DM particle produces more spherical shapes in the particle distribution of simulated halos. While this result is consistent and halos in SIDM simulations systematically exhibit rounder shapes, the actual quantitative differences are relatively small. This can be seen in the upper panel of Fig. \ref{fig:comparison} in which the differences of the semi-axis ratios of the particle distribution between CDM and SIDM simulations are lower than $5\%$. Due to the combination of the stacking power and the fact that the \textit{shear} is sensitive to the whole inner distribution where the impact of considering a non-zero $\sigma/m$ is expected to be larger, weak-lensing estimates show significant differences between the fitted shape parameters from both simulations. %\textit{Moreover they also predict variations in the elongation of the neighbouring mass distributions, where the differences between the simulations are expected to be negligible. }

In the bottom panel of Fig. \ref{fig:comparison} we show the results when accounting for a radial variation of the elongation. In this case, the results based on the stacked particle distribution, show systematically lower values of $q_0$ for the halos in the CDM simulation, but still in agreement within a $5\%$. On the other hand, as discussed in \ref{subsec:stacked_shapes}, the power law slopes differ by a $\sim20\%$. Regarding the results predicted by the lensing analysis, there are no discernible variations in the fitted $q_0$ and $q_{2h}$ parameters. However, we found significant differences in the fitted power law slope, $\alpha$. Our results show a very subtle radial variation in the shapes of halos from the CDM simulation ($-0.04 \lesssim \alpha \lesssim 0.0$), whereas halos in the SIDM simulation exhibit a steeper relationship ($\alpha \lesssim -0.06$). 

Further visualisation of these results is presented in Fig. \ref{fig:result}, in which we show the predicted $q(r)$ relation for the main halo component, according to the fitted parameters, $\alpha$ and $q_0$, together with the fitted elongation for the neighbouring mass contribution, $q_{2h}$. In this Figure, it can be clearly seen how this modelling captures the expected variation of the elongation, in which at the inner radial ranges the halos in the SIDM simulation are rounder towards the centre compared to those in the CDM simulation. For all the samples in both simulations, the neighbouring mass distribution is rounder than the mass distribution of the main halo, except for the \textit{Clustered} sample. This result indicates that the larger clustering amplitude expected for these halos, preferentially occurs in a direction aligned with the halo main orientation. We marked in darker colours the regions in which the main and the 2-halo component are expected to be more dominant, setting the limit at the mean virial radius of the stacked halos. At these regions, the semi-axis ratios of the main halo component from the SIDM and CDM simulations, converge roughly at the same value for most of the samples. However, \textit{Clustered} and \textit{Lower-mass} halo samples show higher differences within this region, being the halos in the SIDM simulation more elongated.

In order to interpret these results, we extend the analysis presented in \ref{subsec:stacked_shapes} by considering all the dark matter particles used to calculate the lensing maps, as opposed to solely the bound particles, and computing the shapes up to projected radii of $5h^{-1}$Mpc. We show in Fig. \ref{fig:ratio_DM} the ratio between the estimated shape parameters from the CDM and the SIDM simulations. When considering only the bound particles the differences between the simulated data are more significant at the inner regions and then remains constant given that we are far from the halo radial extension. On the other hand, the whole particle distribution shows a stepper variation and the semi-axis ratios converge at larger radii. \textit{Clustered} and \textit{Lower-mass} halo sub-sets show a similar behaviour as the modelled from the quadrupole profiles, in which the halos in the SIDM simulation show a more elongated distribution at the outskirts. This last result does not necessarily indicate that these halos are indeed more elongated at the outskirts. Instead it suggests that the neighbouring mass distribution is better aligned with the halo. We should keep in mind that the orientations considered to align the halos and perform the stacking are computed using only the bound dark-matter particles.  This result is in agreement with the presented in \citet{Banerjee2020}. In this work, they obtain from the phase-space of the particles in CDM and SIDM simulations, that the potential becomes more isotropic in the presence of velocity-independent self-interactions.

The observed shapes of the whole DM particle distribution clarify the obtained results presented in \ref{subsec:fitted} and \ref{subsec:radial_var} and summarised in Fig. \ref{fig:comparison} and Fig. \ref{fig:result}. When considering a radial variation, the fitted slopes of the power law are higher for the SIDM since the whole particle distribution shows a steeper radial variation than the predicted according only to the bound particles. On the other hand, results presented in \ref{subsec:fitted}, in which $q_{2h}$ show significant differences between both simulations, is an expected result given that the model tends to compensate for the radial variation mainly present in the SIDM halos. In fact, fitted $q_{2h}$ in the CDM simulation are in agreement for both models within a $\sim 3\%$, while the fitted values for the SIDM are $10\%$ larger when considering a radial variation. 

\section{Summary and conclusions}
\label{sec:conclussion}
In this study, we provide a comprehensive analysis of the application of a stacked weak-lensing approach to determine the shapes of cluster halos, with the aim of gathering insights into the nature of dark matter particles. With this aim, we produce two sets of simulated data, with different cross-sections for the dark-matter particle, $\sigma/m$, but with the same initial conditions. One of the simulated data considers a non-interactive DM particle, $\sigma/m = 0 \,cm^2/g$ (CDM simulation), while the other allows for self-interactions with $\sigma/m = 1\,cm^2/g $ (SIDM simulation). In agreement with previous studies, derived shapes based on the particle distribution of the halos identified in the SIDM simulation are systematically rounder than those in the CDM simulation. However, assessing these differences through observations certainly constitutes a challenging task. 

With the aim of obtaining the halo shapes to discern between the two simulated scenarios, we propose to measure the mean halo elongation by stacking the expected aligned component of the weak-lensing signal. The approach considered in this work to recover the halo shapes, takes advantage of the stacking power,the sensitivity of the lens effect to the entire inner mass distribution within a projected radius and the projection of the signal in the main direction of the halo elongation. This results in radial profiles that efficiently capture the differences introduced due to the shape variation. In fact, our results show that the obtained fitted parameters show larger discrepancies between the simulated data, than when considering the particle distribution. Hence, the adopted methodology highlights the differences between the two simulated data-sets. 

When applying the stacked lensing analysis, we found that the fitted shapes can vary by roughly a $10\%$ for the halos identified in the SIDM simulation compared to those in the CDM simulation, with the latter tending to be more elongated. 
We have also tested a radial variation modelling that considers a relation between the mass elongation and the projected radial distance. The results show a higher slope of the power law for the shape radial variation for the halos in the SIDM simulation, describing rounder mass distributions towards the centre. The shape parameters roughly agree with the measured in CDM simulation at the mean virial radius of the stacked sample. 

We also notice that all the lensing results are supported by the trends of the measured shapes based on the the whole particle distribution included in the analysis. For both models, with and without a radial dependent shape variation, we systematically obtain a more elongated distribution of the neighbouring mass component for the combined halos in the SIDM simulation. This result is more significant for the lower-mass halos and those in denser local environments. The results obtained agree with those presented in \citet{Banerjee2020} where a more coherent infall from larger scales is expected for the halos when considering self-interacting particles.

It is important to take into account that the results derived in this work are bound to the fact that we neglect the impact of the baryon physics in shaping the halos. Nevertheless, the fitted radial ranges include information well beyond the halo radial extension, where the impact of baryons is expected to be less significant. In fact, simulations that include baryon physics found that shape differences in the DM distributions persisted out to large radii and that the shape differences between CDM and SIDM were significantly larger than those between DM-only and hydrodynamical simulations \citep{Robertson2019}. Another effect that will be important to take into account is the estimate of the main-cluster orientation. In the analysis presented we use the particle distribution in order to align the halos for the stacking. These orientations are in principle unknown in observational studies, which consider luminous proxies such as the galaxy member distribution in order to constrain them. Although the member distribution has been shown to offer a good prediction for the halo main orientation \citep{Gonzalez2021b}, hydrodynamical simulations will be necessary to test the impact of this misalignment in the analysis and to account for possible systematic effects. 

Although other effects need to be assessed to improve the calibration of the presented methodology, its application shows a strong potential in capturing the differences introduced by dark-matter particle candidates.
In view of the new upcoming large-scale surveys, such as the Legacy Survey of Space and Time  \citep[LSST,][]{Ivezic2019} and Euclid \citep{Laureijs2011}, that will provide high-quality measurements of lensing statistics, our findings show lensing techniques as a promising approach in order to test the nature of the dark-matter particle. 

\section*{Acknowledgements}

The authors thank M. Boylan-Kolchin and James Bullock for sharing the SIDM implementation, as well as the entire GIZMO team in general, for making the code publicly available.
This work used computational resources from CCAD – Universidad Nacional de Córdoba (\url{https://ccad.unc.edu.ar/}), which are part of SNCAD – MinCyT, República Argentina.
This work was also partially supported by Agencia Nacional de Promoción Científica y Tecnológica
(PICT-2020-SERIEA-01404), the Consejo Nacional de Investigaciones Científicas y Técnicas (CONICET, Argentina) and the Secretaría de Ciencia y Tecnología de la Universidad Nacional de Córdoba
(SeCyT-UNC, Argentina). Also, it was supported by Ministerio de Ciencia e Innovación, Agencia Estatal de Investigación and FEDER (PID2021-123012NA-C44). This work has been supported by the call for grants for Scientific and Technical Equipment 2021 of the State Program for Knowledge Generation and Scientific and Technological Strengthening of the R+D+i System, financed by MCIN/AEI/ 10.13039/501100011033 and the EU NextGeneration/PRTR (Hadoop Cluster for the comprehensive management of massive scientific data, reference EQC2021-007479-P)

%%%%%%%%%%%%%%%%%%%%%%%%%%%%%%%%%%%%%%%%%%%%%%%%%%
\section*{Data Availability}

Simulations and data underlying this article will be shared on reasonable request to the corresponding author.

%%%%%%%%%%%%%%%%%%%% REFERENCES %%%%%%%%%%%%%%%%%%

% The best way to enter references is to use BibTeX:

\bibliographystyle{mnras}
\bibliography{example} % if your bibtex file is called example.bib

\begin{thebibliography}{}
\makeatletter
\relax
\def\mn@urlcharsother{\let\do\@makeother \do\$\do\&\do\#\do\^\do\_\do\%\do\~}
\def\mn@doi{\begingroup\mn@urlcharsother \@ifnextchar [ {\mn@doi@} {\mn@doi@[]}}
\def\mn@doi@[#1]#2{\def\@tempa{#1}\ifx\@tempa\@empty \href {http://dx.doi.org/#2} {doi:#2}\else \href {http://dx.doi.org/#2} {#1}\fi \endgroup}
\def\mn@eprint#1#2{\mn@eprint@#1:#2::\@nil}
\def\mn@eprint@arXiv#1{\href {http://arxiv.org/abs/#1} {{\tt arXiv:#1}}}
\def\mn@eprint@dblp#1{\href {http://dblp.uni-trier.de/rec/bibtex/#1.xml} {dblp:#1}}
\def\mn@eprint@#1:#2:#3:#4\@nil{\def\@tempa {#1}\def\@tempb {#2}\def\@tempc {#3}\ifx \@tempc \@empty \let \@tempc \@tempb \let \@tempb \@tempa \fi \ifx \@tempb \@empty \def\@tempb {arXiv}\fi \@ifundefined {mn@eprint@\@tempb}{\@tempb:\@tempc}{\expandafter \expandafter \csname mn@eprint@\@tempb\endcsname \expandafter{\@tempc}}}

\bibitem[\protect\citeauthoryear{{Alam} et~al.,}{{Alam} et~al.}{2017}]{Alam2017}
{Alam} S.,  et~al., 2017, \mn@doi [\mnras] {10.1093/mnras/stx721}, \href {https://ui.adsabs.harvard.edu/abs/2017MNRAS.470.2617A} {470, 2617}

\bibitem[\protect\citeauthoryear{{Banerjee}, {Adhikari}, {Dalal}, {More}  \& {Kravtsov}}{{Banerjee} et~al.}{2020}]{Banerjee2020}
{Banerjee} A.,  {Adhikari} S.,  {Dalal} N.,  {More} S.,   {Kravtsov} A.,  2020, \mn@doi [\jcap] {10.1088/1475-7516/2020/02/024}, \href {https://ui.adsabs.harvard.edu/abs/2020JCAP...02..024B} {2020, 024}

\bibitem[\protect\citeauthoryear{{Behroozi}, {Wechsler}  \& {Wu}}{{Behroozi} et~al.}{2013}]{Behroozi2013}
{Behroozi} P.~S.,  {Wechsler} R.~H.,   {Wu} H.-Y.,  2013, \mn@doi [\apj] {10.1088/0004-637X/762/2/109}, \href {https://ui.adsabs.harvard.edu/abs/2013ApJ...762..109B} {762, 109}

\bibitem[\protect\citeauthoryear{{Bhattacharya}, {Habib}, {Heitmann}  \& {Vikhlinin}}{{Bhattacharya} et~al.}{2013}]{Bhattacharya2013}
{Bhattacharya} S.,  {Habib} S.,  {Heitmann} K.,   {Vikhlinin} A.,  2013, \mn@doi [\apj] {10.1088/0004-637X/766/1/32}, \href {https://ui.adsabs.harvard.edu/abs/2013ApJ...766...32B} {766, 32}

\bibitem[\protect\citeauthoryear{{Brinckmann}, {Zavala}, {Rapetti}, {Hansen}  \& {Vogelsberger}}{{Brinckmann} et~al.}{2018}]{Brinckmann2018}
{Brinckmann} T.,  {Zavala} J.,  {Rapetti} D.,  {Hansen} S.~H.,   {Vogelsberger} M.,  2018, \mn@doi [\mnras] {10.1093/mnras/stx2782}, \href {https://ui.adsabs.harvard.edu/abs/2018MNRAS.474..746B} {474, 746}

\bibitem[\protect\citeauthoryear{{Bullock} \& {Boylan-Kolchin}}{{Bullock} \& {Boylan-Kolchin}}{2017}]{Bullock2017}
{Bullock} J.~S.,  {Boylan-Kolchin} M.,  2017, \mn@doi [\araa] {10.1146/annurev-astro-091916-055313}, \href {https://ui.adsabs.harvard.edu/abs/2017ARA&A..55..343B} {55, 343}

\bibitem[\protect\citeauthoryear{{Bullock}, {Kolatt}, {Sigad}, {Somerville}, {Kravtsov}, {Klypin}, {Primack}  \& {Dekel}}{{Bullock} et~al.}{2001}]{Bullock2001}
{Bullock} J.~S.,  {Kolatt} T.~S.,  {Sigad} Y.,  {Somerville} R.~S.,  {Kravtsov} A.~V.,  {Klypin} A.~A.,  {Primack} J.~R.,   {Dekel} A.,  2001, \mn@doi [\mnras] {10.1046/j.1365-8711.2001.04068.x}, \href {https://ui.adsabs.harvard.edu/abs/2001MNRAS.321..559B} {321, 559}

\bibitem[\protect\citeauthoryear{{Ching} et~al.,}{{Ching} et~al.}{2017}]{Ching2017}
{Ching} J.~H.~Y.,  et~al., 2017, \mn@doi [\mnras] {10.1093/mnras/stx1173}, \href {https://ui.adsabs.harvard.edu/abs/2017MNRAS.469.4584C} {469, 4584}

\bibitem[\protect\citeauthoryear{{Clampitt} \& {Jain}}{{Clampitt} \& {Jain}}{2016}]{Clampitt2016}
{Clampitt} J.,  {Jain} B.,  2016, \mn@doi [\mnras] {10.1093/mnras/stw254}, \href {https://ui.adsabs.harvard.edu/abs/2016MNRAS.457.4135C} {457, 4135}

\bibitem[\protect\citeauthoryear{{Despali}, {Walls}, {Vegetti}, {Sparre}, {Vogelsberger}  \& {Zavala}}{{Despali} et~al.}{2022}]{Despali2022}
{Despali} G.,  {Walls} L.~G.,  {Vegetti} S.,  {Sparre} M.,  {Vogelsberger} M.,   {Zavala} J.,  2022, \mn@doi [\mnras] {10.1093/mnras/stac2521}, \href {https://ui.adsabs.harvard.edu/abs/2022MNRAS.516.4543D} {516, 4543}

\bibitem[\protect\citeauthoryear{{Diemer}}{{Diemer}}{2018}]{Diemer2018}
{Diemer} B.,  2018, \mn@doi [\apjs] {10.3847/1538-4365/aaee8c}, \href {https://ui.adsabs.harvard.edu/abs/2018ApJS..239...35D} {239, 35}

\bibitem[\protect\citeauthoryear{{Dubinski} \& {Carlberg}}{{Dubinski} \& {Carlberg}}{1991}]{DubinskiCarlberg1991}
{Dubinski} J.,  {Carlberg} R.~G.,  1991, \mn@doi [\apj] {10.1086/170451}, \href {https://ui.adsabs.harvard.edu/abs/1991ApJ...378..496D} {378, 496}

\bibitem[\protect\citeauthoryear{{Duffy}, {Schaye}, {Kay}  \& {Dalla Vecchia}}{{Duffy} et~al.}{2008}]{Duffy2008}
{Duffy} A.~R.,  {Schaye} J.,  {Kay} S.~T.,   {Dalla Vecchia} C.,  2008, \mn@doi [\mnras] {10.1111/j.1745-3933.2008.00537.x}, \href {https://ui.adsabs.harvard.edu/abs/2008MNRAS.390L..64D} {390, L64}

\bibitem[\protect\citeauthoryear{{Einasto} \& {Haud}}{{Einasto} \& {Haud}}{1989}]{Einasto1989}
{Einasto} J.,  {Haud} U.,  1989, \aap, \href {https://ui.adsabs.harvard.edu/abs/1989A&A...223...89E} {223, 89}

\bibitem[\protect\citeauthoryear{{Evans} \& {Bridle}}{{Evans} \& {Bridle}}{2009}]{Evans2009}
{Evans} A. K.~D.,  {Bridle} S.,  2009, \mn@doi [\apj] {10.1088/0004-637X/695/2/1446}, \href {https://ui.adsabs.harvard.edu/abs/2009ApJ...695.1446E} {695, 1446}

\bibitem[\protect\citeauthoryear{{Flores} \& {Primack}}{{Flores} \& {Primack}}{1994}]{Flores1994}
{Flores} R.~A.,  {Primack} J.~R.,  1994, \mn@doi [\apjl] {10.1086/187350}, \href {https://ui.adsabs.harvard.edu/abs/1994ApJ...427L...1F} {427, L1}

\bibitem[\protect\citeauthoryear{{Foreman-Mackey}, {Hogg}, {Lang}  \& {Goodman}}{{Foreman-Mackey} et~al.}{2013}]{Foreman2013}
{Foreman-Mackey} D.,  {Hogg} D.~W.,  {Lang} D.,   {Goodman} J.,  2013, \mn@doi [\pasp] {10.1086/670067}, \href {https://ui.adsabs.harvard.edu/abs/2013PASP..125..306F} {125, 306}

\bibitem[\protect\citeauthoryear{{Gonzalez}, {Makler}, {Garc{\'\i}a Lambas}, {Chalela}, {Pereira}, {Van Waerbeke}, {Shan}  \& {Erben}}{{Gonzalez} et~al.}{2021a}]{Gonzalez2021}
{Gonzalez} E.~J.,  {Makler} M.,  {Garc{\'\i}a Lambas} D.,  {Chalela} M.,  {Pereira} M. E.~S.,  {Van Waerbeke} L.,  {Shan} H.,   {Erben} T.,  2021a, \mn@doi [\mnras] {10.1093/mnras/staa3570}, \href {https://ui.adsabs.harvard.edu/abs/2021MNRAS.501.5239G} {501, 5239}

\bibitem[\protect\citeauthoryear{{Gonzalez}, {Ragone-Figueroa}, {Donzelli}, {Makler}, {Garc{\'\i}a Lambas}  \& {Granato}}{{Gonzalez} et~al.}{2021b}]{Gonzalez2021b}
{Gonzalez} E.~J.,  {Ragone-Figueroa} C.,  {Donzelli} C.~J.,  {Makler} M.,  {Garc{\'\i}a Lambas} D.,   {Granato} G.~L.,  2021b, \mn@doi [\mnras] {10.1093/mnras/stab2585}, \href {https://ui.adsabs.harvard.edu/abs/2021MNRAS.508.1280G} {508, 1280}

\bibitem[\protect\citeauthoryear{{Gonzalez}, {Hoffmann}, {Gazta{\~n}aga}, {Garc{\'\i}a Lambas}, {Fosalba}, {Crocce}, {Castander}  \& {Makler}}{{Gonzalez} et~al.}{2022}]{Gonzalez2022}
{Gonzalez} E.~J.,  {Hoffmann} K.,  {Gazta{\~n}aga} E.,  {Garc{\'\i}a Lambas} D.~R.,  {Fosalba} P.,  {Crocce} M.,  {Castander} F.~J.,   {Makler} M.,  2022, \mn@doi [\mnras] {10.1093/mnras/stac3038}, \href {https://ui.adsabs.harvard.edu/abs/2022MNRAS.517.4827G} {517, 4827}

\bibitem[\protect\citeauthoryear{{Hahn} \& {Abel}}{{Hahn} \& {Abel}}{2011}]{Hahn2011}
{Hahn} O.,  {Abel} T.,  2011, \mn@doi [\mnras] {10.1111/j.1365-2966.2011.18820.x}, \href {https://ui.adsabs.harvard.edu/abs/2011MNRAS.415.2101H} {415, 2101}

\bibitem[\protect\citeauthoryear{{Hopkins}}{{Hopkins}}{2015}]{Hopkins2015}
{Hopkins} P.~F.,  2015, \mn@doi [\mnras] {10.1093/mnras/stv195}, \href {https://ui.adsabs.harvard.edu/abs/2015MNRAS.450...53H} {450, 53}

\bibitem[\protect\citeauthoryear{{Huang}, {Mandelbaum}, {Freeman}, {Chen}, {Rozo}, {Rykoff}  \& {Baxter}}{{Huang} et~al.}{2016}]{huang2016}
{Huang} H.-J.,  {Mandelbaum} R.,  {Freeman} P.~E.,  {Chen} Y.-C.,  {Rozo} E.,  {Rykoff} E.,   {Baxter} E.~J.,  2016, \mn@doi [\mnras] {10.1093/mnras/stw1982}, \href {http://adsabs.harvard.edu/abs/2016MNRAS.463..222H} {463, 222}

\bibitem[\protect\citeauthoryear{{Ishiyama} et~al.,}{{Ishiyama} et~al.}{2021}]{Ishiyama2021}
{Ishiyama} T.,  et~al., 2021, \mn@doi [\mnras] {10.1093/mnras/stab1755}, \href {https://ui.adsabs.harvard.edu/abs/2021MNRAS.506.4210I} {506, 4210}

\bibitem[\protect\citeauthoryear{{Ivezi{\'c}} et~al.,}{{Ivezi{\'c}} et~al.}{2019}]{Ivezic2019}
{Ivezi{\'c}} {\v{Z}}.,  et~al., 2019, \mn@doi [\apj] {10.3847/1538-4357/ab042c}, \href {https://ui.adsabs.harvard.edu/abs/2019ApJ...873..111I} {873, 111}

\bibitem[\protect\citeauthoryear{{Jee}, {Hoekstra}, {Mahdavi}  \& {Babul}}{{Jee} et~al.}{2014}]{Jee2014}
{Jee} M.~J.,  {Hoekstra} H.,  {Mahdavi} A.,   {Babul} A.,  2014, \mn@doi [\apj] {10.1088/0004-637X/783/2/78}, \href {https://ui.adsabs.harvard.edu/abs/2014ApJ...783...78J} {783, 78}

\bibitem[\protect\citeauthoryear{{Kaiser} \& {Squires}}{{Kaiser} \& {Squires}}{1993}]{Kaiser1993}
{Kaiser} N.,  {Squires} G.,  1993, \mn@doi [\apj] {10.1086/172297}, \href {https://ui.adsabs.harvard.edu/abs/1993ApJ...404..441K} {404, 441}

\bibitem[\protect\citeauthoryear{{Katz}}{{Katz}}{1991}]{Katz1991}
{Katz} N.,  1991, \mn@doi [\apj] {10.1086/169696}, \href {https://ui.adsabs.harvard.edu/abs/1991ApJ...368..325K} {368, 325}

\bibitem[\protect\citeauthoryear{{Lackner} \& {Gunn}}{{Lackner} \& {Gunn}}{2012}]{Lackner2012}
{Lackner} C.,  {Gunn} J.,  2012, in American Astronomical Society Meeting Abstracts \#219. p. 417.02

\bibitem[\protect\citeauthoryear{{Laureijs} et~al.,}{{Laureijs} et~al.}{2011}]{Laureijs2011}
{Laureijs} R.,  et~al., 2011, arXiv e-prints, \href {https://ui.adsabs.harvard.edu/abs/2011arXiv1110.3193L} {p. arXiv:1110.3193}

\bibitem[\protect\citeauthoryear{{McClintock} et~al.,}{{McClintock} et~al.}{2019}]{McClintock2019}
{McClintock} T.,  et~al., 2019, \mn@doi [\mnras] {10.1093/mnras/sty2711}, \href {https://ui.adsabs.harvard.edu/abs/2019MNRAS.482.1352M} {482, 1352}

\bibitem[\protect\citeauthoryear{{Meskhidze}, {Mercado}, {Sameie}, {Robles}, {Bullock}, {Kaplinghat}  \& {Weatherall}}{{Meskhidze} et~al.}{2022}]{Meskhidze2022}
{Meskhidze} H.,  {Mercado} F.~J.,  {Sameie} O.,  {Robles} V.~H.,  {Bullock} J.~S.,  {Kaplinghat} M.,   {Weatherall} J.~O.,  2022, \mn@doi [\mnras] {10.1093/mnras/stac1056}, \href {https://ui.adsabs.harvard.edu/abs/2022MNRAS.513.2600M} {513, 2600}

\bibitem[\protect\citeauthoryear{{Miralda-Escud{\'e}}}{{Miralda-Escud{\'e}}}{2002}]{Miralda-escude2002}
{Miralda-Escud{\'e}} J.,  2002, \mn@doi [\apj] {10.1086/324138}, \href {https://ui.adsabs.harvard.edu/abs/2002ApJ...564...60M} {564, 60}

\bibitem[\protect\citeauthoryear{{Muldrew} et~al.,}{{Muldrew} et~al.}{2012}]{Muldrew2012}
{Muldrew} S.~I.,  et~al., 2012, \mn@doi [\mnras] {10.1111/j.1365-2966.2011.19922.x}, \href {https://ui.adsabs.harvard.edu/abs/2012MNRAS.419.2670M} {419, 2670}

\bibitem[\protect\citeauthoryear{{Navarro}, {Frenk}  \& {White}}{{Navarro} et~al.}{1997}]{Navarro97}
{Navarro} J.~F.,  {Frenk} C.~S.,   {White} S. D.~M.,  1997, \mn@doi [\apj] {10.1086/304888}, \href {https://ui.adsabs.harvard.edu/abs/1997ApJ...490..493N} {490, 493}

\bibitem[\protect\citeauthoryear{{Neto} et~al.,}{{Neto} et~al.}{2007}]{Neto2007}
{Neto} A.~F.,  et~al., 2007, \mn@doi [\mnras] {10.1111/j.1365-2966.2007.12381.x}, \href {https://ui.adsabs.harvard.edu/abs/2007MNRAS.381.1450N} {381, 1450}

\bibitem[\protect\citeauthoryear{{Newman}, {Treu}, {Ellis}, {Sand}, {Richard}, {Marshall}, {Capak}  \& {Miyazaki}}{{Newman} et~al.}{2009}]{Newman2009}
{Newman} A.~B.,  {Treu} T.,  {Ellis} R.~S.,  {Sand} D.~J.,  {Richard} J.,  {Marshall} P.~J.,  {Capak} P.,   {Miyazaki} S.,  2009, \mn@doi [\apj] {10.1088/0004-637X/706/2/1078}, \href {https://ui.adsabs.harvard.edu/abs/2009ApJ...706.1078N} {706, 1078}

\bibitem[\protect\citeauthoryear{{Niemiec} et~al.,}{{Niemiec} et~al.}{2017}]{Niemiec2017}
{Niemiec} A.,  et~al., 2017, \mn@doi [\mnras] {10.1093/mnras/stx1667}, \href {https://ui.adsabs.harvard.edu/abs/2017MNRAS.471.1153N} {471, 1153}

\bibitem[\protect\citeauthoryear{{Oguri}, {Takada}, {Okabe}  \& {Smith}}{{Oguri} et~al.}{2010}]{Oguri2010}
{Oguri} M.,  {Takada} M.,  {Okabe} N.,   {Smith} G.~P.,  2010, \mn@doi [\mnras] {10.1111/j.1365-2966.2010.16622.x}, \href {https://ui.adsabs.harvard.edu/abs/2010MNRAS.405.2215O} {405, 2215}

\bibitem[\protect\citeauthoryear{{Okoli}}{{Okoli}}{2017}]{Okoli2017}
{Okoli} C.,  2017, arXiv e-prints, \href {https://ui.adsabs.harvard.edu/abs/2017arXiv171105277O} {p. arXiv:1711.05277}

\bibitem[\protect\citeauthoryear{{Pereira} et~al.,}{{Pereira} et~al.}{2020}]{Pereira2020}
{Pereira} M.~E.~S.,  et~al., 2020, \mn@doi [\mnras] {10.1093/mnras/staa2687}, \href {https://ui.adsabs.harvard.edu/abs/2020MNRAS.498.5450P} {498, 5450}

\bibitem[\protect\citeauthoryear{{Peter}, {Rocha}, {Bullock}  \& {Kaplinghat}}{{Peter} et~al.}{2013}]{Peter2013}
{Peter} A. H.~G.,  {Rocha} M.,  {Bullock} J.~S.,   {Kaplinghat} M.,  2013, \mn@doi [\mnras] {10.1093/mnras/sts535}, \href {https://ui.adsabs.harvard.edu/abs/2013MNRAS.430..105P} {430, 105}

\bibitem[\protect\citeauthoryear{{Planck Collaboration} et~al.,}{{Planck Collaboration} et~al.}{2020}]{Aghanim2020}
{Planck Collaboration} et~al., 2020, \mn@doi [\aap] {10.1051/0004-6361/201833910}, \href {https://ui.adsabs.harvard.edu/abs/2020A&A...641A...6P} {641, A6}

\bibitem[\protect\citeauthoryear{{Randall}, {Markevitch}, {Clowe}, {Gonzalez}  \& {Brada{\v{c}}}}{{Randall} et~al.}{2008}]{Randall2008}
{Randall} S.~W.,  {Markevitch} M.,  {Clowe} D.,  {Gonzalez} A.~H.,   {Brada{\v{c}}} M.,  2008, \mn@doi [\apj] {10.1086/587859}, \href {https://ui.adsabs.harvard.edu/abs/2008ApJ...679.1173R} {679, 1173}

\bibitem[\protect\citeauthoryear{{Retana-Montenegro}, {van Hese}, {Gentile}, {Baes}  \& {Frutos-Alfaro}}{{Retana-Montenegro} et~al.}{2012}]{Retana2012}
{Retana-Montenegro} E.,  {van Hese} E.,  {Gentile} G.,  {Baes} M.,   {Frutos-Alfaro} F.,  2012, \mn@doi [\aap] {10.1051/0004-6361/201118543}, \href {https://ui.adsabs.harvard.edu/abs/2012A&A...540A..70R} {540, A70}

\bibitem[\protect\citeauthoryear{{Robertson}, {Harvey}, {Massey}, {Eke}, {McCarthy}, {Jauzac}, {Li}  \& {Schaye}}{{Robertson} et~al.}{2019}]{Robertson2019}
{Robertson} A.,  {Harvey} D.,  {Massey} R.,  {Eke} V.,  {McCarthy} I.~G.,  {Jauzac} M.,  {Li} B.,   {Schaye} J.,  2019, \mn@doi [\mnras] {10.1093/mnras/stz1815}, \href {https://ui.adsabs.harvard.edu/abs/2019MNRAS.488.3646R} {488, 3646}

\bibitem[\protect\citeauthoryear{{Robertson}, {Huff}  \& {Markovi{\v{c}}}}{{Robertson} et~al.}{2023}]{Robertson2023}
{Robertson} A.,  {Huff} E.,   {Markovi{\v{c}}} K.,  2023, \mn@doi [\mnras] {10.1093/mnras/stad655}, \href {https://ui.adsabs.harvard.edu/abs/2023MNRAS.521.3172R} {521, 3172}

\bibitem[\protect\citeauthoryear{{Rocha}, {Peter}, {Bullock}, {Kaplinghat}, {Garrison-Kimmel}, {O{\~n}orbe}  \& {Moustakas}}{{Rocha} et~al.}{2013}]{Rocha2013}
{Rocha} M.,  {Peter} A. H.~G.,  {Bullock} J.~S.,  {Kaplinghat} M.,  {Garrison-Kimmel} S.,  {O{\~n}orbe} J.,   {Moustakas} L.~A.,  2013, \mn@doi [\mnras] {10.1093/mnras/sts514}, \href {https://ui.adsabs.harvard.edu/abs/2013MNRAS.430...81R} {430, 81}

\bibitem[\protect\citeauthoryear{{Rubin} \& {Ford}}{{Rubin} \& {Ford}}{1970}]{Rubin1970}
{Rubin} V.~C.,  {Ford} W.~Kent J.,  1970, \mn@doi [\apj] {10.1086/150317}, \href {https://ui.adsabs.harvard.edu/abs/1970ApJ...159..379R} {159, 379}

\bibitem[\protect\citeauthoryear{{Sand}, {Treu}, {Smith}  \& {Ellis}}{{Sand} et~al.}{2004}]{Sand2004}
{Sand} D.~J.,  {Treu} T.,  {Smith} G.~P.,   {Ellis} R.~S.,  2004, \mn@doi [\apj] {10.1086/382146}, \href {https://ui.adsabs.harvard.edu/abs/2004ApJ...604...88S} {604, 88}

\bibitem[\protect\citeauthoryear{{Santucci} et~al.,}{{Santucci} et~al.}{2023}]{Santucci2023}
{Santucci} G.,  et~al., 2023, \mn@doi [\mnras] {10.1093/mnras/stad713}, \href {https://ui.adsabs.harvard.edu/abs/2023MNRAS.521.2671S} {521, 2671}

\bibitem[\protect\citeauthoryear{{Schneider} \& {Weiss}}{{Schneider} \& {Weiss}}{1991}]{Schneider1991}
{Schneider} P.,  {Weiss} A.,  1991, \aap, \href {https://ui.adsabs.harvard.edu/abs/1991A&A...247..269S} {247, 269}

\bibitem[\protect\citeauthoryear{{Seljak} \& {Warren}}{{Seljak} \& {Warren}}{2004}]{Seljak2004}
{Seljak} U.,  {Warren} M.~S.,  2004, \mn@doi [\mnras] {10.1111/j.1365-2966.2004.08297.x}, \href {https://ui.adsabs.harvard.edu/abs/2004MNRAS.355..129S} {355, 129}

\bibitem[\protect\citeauthoryear{{Shin}, {Clampitt}, {Jain}, {Bernstein}, {Neil}, {Rozo}  \& {Rykoff}}{{Shin} et~al.}{2018}]{shin2018}
{Shin} T.-h.,  {Clampitt} J.,  {Jain} B.,  {Bernstein} G.,  {Neil} A.,  {Rozo} E.,   {Rykoff} E.,  2018, \mn@doi [\mnras] {10.1093/mnras/stx3366}, \href {http://adsabs.harvard.edu/abs/2018MNRAS.475.2421S} {475, 2421}

\bibitem[\protect\citeauthoryear{{Simon}, {Bolatto}, {Leroy}, {Blitz}  \& {Gates}}{{Simon} et~al.}{2005}]{Simon2005}
{Simon} J.~D.,  {Bolatto} A.~D.,  {Leroy} A.,  {Blitz} L.,   {Gates} E.~L.,  2005, \mn@doi [\apj] {10.1086/427684}, \href {https://ui.adsabs.harvard.edu/abs/2005ApJ...621..757S} {621, 757}

\bibitem[\protect\citeauthoryear{{Spergel} \& {Steinhardt}}{{Spergel} \& {Steinhardt}}{2000}]{Spergel2000}
{Spergel} D.~N.,  {Steinhardt} P.~J.,  2000, \mn@doi [\prl] {10.1103/PhysRevLett.84.3760}, \href {https://ui.adsabs.harvard.edu/abs/2000PhRvL..84.3760S} {84, 3760}

\bibitem[\protect\citeauthoryear{{Tinker}, {Robertson}, {Kravtsov}, {Klypin}, {Warren}, {Yepes}  \& {Gottl{\"o}ber}}{{Tinker} et~al.}{2010}]{Tinker2010}
{Tinker} J.~L.,  {Robertson} B.~E.,  {Kravtsov} A.~V.,  {Klypin} A.,  {Warren} M.~S.,  {Yepes} G.,   {Gottl{\"o}ber} S.,  2010, \mn@doi [\apj] {10.1088/0004-637X/724/2/878}, \href {https://ui.adsabs.harvard.edu/abs/2010ApJ...724..878T} {724, 878}

\bibitem[\protect\citeauthoryear{{Tulin} \& {Yu}}{{Tulin} \& {Yu}}{2018}]{Tulin2018}
{Tulin} S.,  {Yu} H.-B.,  2018, \mn@doi [\physrep] {10.1016/j.physrep.2017.11.004}, \href {https://ui.adsabs.harvard.edu/abs/2018PhR...730....1T} {730, 1}

\bibitem[\protect\citeauthoryear{{Umetsu} et~al.,}{{Umetsu} et~al.}{2012}]{Umetsu2012}
{Umetsu} K.,  et~al., 2012, \mn@doi [\apj] {10.1088/0004-637X/755/1/56}, \href {https://ui.adsabs.harvard.edu/abs/2012ApJ...755...56U} {755, 56}

\bibitem[\protect\citeauthoryear{{Walker} \& {Pe{\~n}arrubia}}{{Walker} \& {Pe{\~n}arrubia}}{2011}]{Walker2011}
{Walker} M.~G.,  {Pe{\~n}arrubia} J.,  2011, \mn@doi [\apj] {10.1088/0004-637X/742/1/20}, \href {https://ui.adsabs.harvard.edu/abs/2011ApJ...742...20W} {742, 20}

\bibitem[\protect\citeauthoryear{{Weinberg}, {Mortonson}, {Eisenstein}, {Hirata}, {Riess}  \& {Rozo}}{{Weinberg} et~al.}{2013}]{Weinberg2013}
{Weinberg} D.~H.,  {Mortonson} M.~J.,  {Eisenstein} D.~J.,  {Hirata} C.,  {Riess} A.~G.,   {Rozo} E.,  2013, \mn@doi [\physrep] {10.1016/j.physrep.2013.05.001}, \href {https://ui.adsabs.harvard.edu/abs/2013PhR...530...87W} {530, 87}

\bibitem[\protect\citeauthoryear{{Weinberg}, {Bullock}, {Governato}, {Kuzio de Naray}  \& {Peter}}{{Weinberg} et~al.}{2015}]{Weinberg2015}
{Weinberg} D.~H.,  {Bullock} J.~S.,  {Governato} F.,  {Kuzio de Naray} R.,   {Peter} A. H.~G.,  2015, \mn@doi [Proceedings of the National Academy of Science] {10.1073/pnas.1308716112}, \href {https://ui.adsabs.harvard.edu/abs/2015PNAS..11212249W} {112, 12249}

\bibitem[\protect\citeauthoryear{{Wittman}, {Golovich}  \& {Dawson}}{{Wittman} et~al.}{2018}]{Wittman2018}
{Wittman} D.,  {Golovich} N.,   {Dawson} W.~A.,  2018, \mn@doi [\apj] {10.3847/1538-4357/aaee77}, \href {https://ui.adsabs.harvard.edu/abs/2018ApJ...869..104W} {869, 104}

\bibitem[\protect\citeauthoryear{{Zemp}, {Gnedin}, {Gnedin}  \& {Kravtsov}}{{Zemp} et~al.}{2011}]{Zemp2011}
{Zemp} M.,  {Gnedin} O.~Y.,  {Gnedin} N.~Y.,   {Kravtsov} A.~V.,  2011, \mn@doi [\apjs] {10.1088/0067-0049/197/2/30}, \href {https://ui.adsabs.harvard.edu/abs/2011ApJS..197...30Z} {197, 30}

\bibitem[\protect\citeauthoryear{{van Uitert} et~al.,}{{van Uitert} et~al.}{2017}]{Uitert2017}
{van Uitert} E.,  et~al., 2017, \mn@doi [\mnras] {10.1093/mnras/stx344}, \href {https://ui.adsabs.harvard.edu/abs/2017MNRAS.467.4131V} {467, 4131}

\makeatother
\end{thebibliography}

% Alternatively you could enter them by hand, like this:
% This method is tedious and prone to error if you have lots of references
%\begin{thebibliography}{99}
%\bibitem[\protect\citeauthoryear{Author}{2012}]{Author2012}
%Author A.~N., 2013, Journal of Improbable Astronomy, 1, 1
%\bibitem[\protect\citeauthoryear{Others}{2013}]{Others2013}
%Others S., 2012, Journal of Interesting Stuff, 17, 198
%\end{thebibliography}

%%%%%%%%%%%%%%%%%%%%%%%%%%%%%%%%%%%%%%%%%%%%%%%%%%

%%%%%%%%%%%%%%%%% APPENDICES %%%%%%%%%%%%%%%%%%%%%

%\appendix

%\section{Some extra material}

%%%%%%%%%%%%%%%%%%%%%%%%%%%%%%%%%%%%%%%%%%%%%%%%%%

% Don't change these lines
\bsp	% typesetting comment
\label{lastpage}
\end{document}